\documentclass{aa}
\usepackage{natbib}
\bibpunct{(}{)}{;}{a}{}{,} 
\input{bib.def}

\usepackage{xcolor,colortbl}
\usepackage{wasysym}
\definecolor{MyRed}{rgb}{0.9,0.0,0.0} 
\definecolor{MyMediumBlue}{rgb}{0.7,0.72,1.0} 
\definecolor{MyGreen}{rgb}{0.0,0.9,0.5} 
\definecolor{SWDarkRed}{rgb}{0.5,0.0,0.0} 
\usepackage[colorlinks=true,urlcolor=blue,citecolor=blue,linkcolor=blue]{hyperref}

\newcommand{\ha}[1]{{#1}}
\newcommand{\hea}[1]{{#1}}
\newcommand{\hab}[1]{{#1}}
\usepackage{graphicx}
\usepackage{txfonts}

\usepackage{ulem}
\usepackage{multirow}
\begin{document} 

   \title{Utilizing the slope of the brightness temperature continuum as a diagnostic tool of solar ALMA observations}

   \author{Henrik Eklund \inst{1,2,3}
   \and 
    Miko\l{}aj Szydlarski \inst{1,2}
    \and
    Sven Wedemeyer \inst{1,2}
  }
  \authorrunning{Eklund {et~al.}}
   \institute{
   Rosseland Centre for Solar  Physics, University of Oslo, Postboks 1029 Blindern, N-0315 Oslo, Norway
            \and
            Institute of  Theoretical Astrophysics, University of Oslo, Postboks 1029 Blindern, N-0315 Oslo, Norway 
            \and
               Institute for Solar Physics, Department of Astronomy, Stockholm University AlbaNova University Centre, SE-106 91 Stockholm, Sweden\\
            \email{henrik.eklund@astro.su.se}
} 
 

   \date{Received --- ; accepted --- }

\abstract{
The intensity of radiation at millimeter wavelengths from the solar atmosphere is closely related to the plasma temperature and the height of formation of the radiation is wavelength dependent.
From that follows that the slope of the intensity continuum, or brightness temperature continuum, samples the local gradient of the gas temperature of the sampled layers in the solar atmosphere.
}
{
We aim to show the added information and diagnostics potential of the solar atmosphere that comes with measuring the slope of the brightness temperature continuum.
}
{
We make use of solar observations from the Atacama Large Millimeter/sub-millimeter Array (ALMA) and perform estimations and prediction \hea{of the slope of the  continuum brightness temperature} based on \hea{differences between} calculations of synthetic observables at wavelengths corresponding to \hea{different sub-bands at the opposite sides of} the ALMA receiver band~3 (\hea{2.8--3.2~mm}) and band~6 (\hea{1.20--1.31~mm}) from a state-of-the-art numerical three-dimensional radiation-magnetohydrodynamic simulation featuring quiet Sun conditions with an enhanced network patch.
}
{
The sign of the brightness temperature slope indicates temperature changes with increasing height at the sampled layers.  A positive sign implies an increase in temperature while a negative sign implies a decrease.

The brightness temperature differences between the sub-bands across the field of view of the simulation typically span from -0.4~kK to 0.75~kK
for band 3 and -0.2~kK to 0.3~kK at band 6. 
The network patches are dominated by large positive \hea{slopes} while the quiet Sun region show a mixture of positive and negative \hea{slopes}. 
As the \hea{slope of the continuum is} coupled to the small-scale dynamics, \hea{a negative slope} is seen typically under quiet Sun conditions as a result of  propagating shock waves and the corresponding post-shock regions. 
The temporal evolution of the \hea{slopes} can therefore be used to identify shocks. 
The observability of the \hea{slope of brightness temperatures} is estimated at bands~3 and 6, for different angular resolutions corresponding to ALMA observations.
\hea{The simulations also show that the intensity of the radiation of both bands~3 and 6 can origin from several major components at different heights simultaneously, which is strongly dependent on the small-scale dynamics and seen in both quiet Sun and network patches.}

In-depth analysis of selected shock waves propagating upwards in the atmosphere shows that the delay of shock signatures between two wavelengths (e.g., bands~6 and 3) 
does not necessarily reflect the propagation speed of the shock front, but \hea{could be caused by different} rate of change of opacity of \hea{above-lying} layers, at these wavelengths.
}
{
The slope of the brightness temperature continuum 
sampled at different 
ALMA receiver sub-bands  
serves as indicator 
of the slope of the local plasma temperature at the sampled heights in the atmosphere, which offers new diagnostic possibilities to measure 
the \hea{underlying physical properties of small scale  dynamic features}, thus contributing to the understanding of such features and the related  transport of \hea{energy and heat} in the chromosphere. 
}


   \maketitle


\section{Introduction}
The continuum radiation at millimetre (mm)  wavelengths  originates from chromospheric heights and forms  under local thermodynamic equilibrium conditions, so that the measured intensities are linked to the temperature of the emitting layers \citep[see e.g.,][and references therein]{2016SSRv..200....1W}.
With the start of regular solar observations with the Atacama Large Millimeter/sub-millimeter Array (ALMA) in 2016, the abilities of observing the Sun at  millimeter wavelengths has taken a leap.
Since then, much work has been put into developing and improving techniques for reconstructing high-cadence time-series from the interferometric measurement sets, which results in successfully producing science ready data. Some of these resulting high-level datasets are publicly available in the Solar ALMA Science Archive \citep[SALSA;][]{2021arXiv210902374H}.  
An increasing number of studies making use of solar ALMA data has been published \citep[see e.g., references within][]{2021arXiv210902374H}.
\ha{In} most of these works, the spectral domain is sacrificed and the flux (and equivalently the brightness temperatures) are reconstructed from the \ha{combined} interferometric data across the whole receiver band in order to increase the signal-to-noise ratio and thus the quality of the resulting images.

The intensities in each of the ALMA receiver bands are sampled in two pairs of sub-bands \citep{ALMA_Tech_Hand_8.3}.
In the current work, we make use of these sub-bands by splitting the data and reconstruct images separately 
\hea{for each of} the individual sub-bands.
ALMA data split into sub-bands was analysed by \cite{2019ApJ...875..163R} in order to estimate the optical thickness of radiation at band~3 (3.0~mm; Table~\ref{tab:almasubbands}) in observations of X-ray bright points, and by
\cite{2019A&A...622A.150J} 
\hea{who made a comparison of brightness temperatures seen at the sub-bands of ALMA band~6 (1.25~mm)
and simultaneous observations at ultraviolet wavelengths with the Interface Region Imaging Spectrograph \citep[IRIS;][]{2014SoPh..289.2733D}, 
at and in the vicinity of a sunspot}. 
\cite{2021RSPTA.37900185E} investigated the diagnostic possibilities of sub-bands based on numerical simulations. The authors 
calculated the synthetic observables \hea{\citep[using the Advanced Radiative Transfer (ART) code][]{art_2021} }at different frequencies of ALMA receiver band~6 (1.20~mm -- 1.31~mm) from a 3D-MHD \textit{Bifrost} \citep{2011A&A...531A.154G} simulation of an enhanced network region with surrounding quiet Sun and analysed the slope of the brightness temperature continuum during the propagation of a shock wave. 
\hea{The authors show that the slope of the brightness temperature continuum can be used for probing and study the small-scale features of the chromosphere.}

In the current work, we further develop several aspects of ALMA sub-bands as a new diagnostic tool.  
We analyse the slope of the continuum brightness temperature of both bands~3 and 6 over the entire FOV of a \textit{Bifrost} simulation \citep{2016A&A...585A...4C}, which shows different characteristics depending on the magnetic field topology, and extend the in-depth analysis of the propagating shock fronts in the quiet Sun region and add a similar in-depth analysis of the network region. 
While the observability of small-scale dynamic brightening events at different angular resolutions and receiver bands of ALMA observations was investigated by \cite{2021A&A...656A.68E}, 
in the current work, we also address the observability of the continuum slope at different angular resolutions corresponding to those of ALMA observations.
In addition to the predictions based on numerical simulations, we also look into observational data of the quiet Sun where we detect shock wave signatures and compare the slopes of the continuum brightness temperature to those from the simulations.

This work is structured in the following way. In Sect.~\ref{sec:methods} the setup of the numerical simulations and methodologies of the radiative transfer calculations and degradation  of the resulting observables for different angular resolutions are described. In Sect.~\ref{sec:results}, the results are presented, including the implications of the continuum slope as derived from the sub-band \hea{differences.
In} addition, the observability of the slope of the continuum at resolutions corresponding to ALMA observations is investigated. The subsequent comparison to observational solar ALMA data shows that the predictions from the simulations are confirmed by the observations.
In Sect.~\ref{sec:disc} we take up the discussion on some topics that are important to keep in mind when performing the analysis of observational data and finally in Sect.~\ref{sec:conc} we conclude the findings.

\section{Methods}\label{sec:methods}

\subsection{Three-dimensional magnetohydrodynamic simulations}\label{sec:method-bifrost_model}

A three-dimensional model of the solar atmosphere from the numerical radiative magnetohydrodynamic code \textit{Bifrost} was used \citep{2011A&A...531A.154G, 2016A&A...585A...4C}. The simulation accounts for non-equilibrium hydrogen ionization. 
The simulation shows some network patches with opposite polarity of the magnetic field approximately $8$~Mm apart, surrounded by quiet-Sun like region. The simulation box has the extent $(x,y,z)=(24,24,17)$~Mm, with the bottom boundary at $2.5$~Mm below the photosphere. In the vertical direction, ($z$), there are $496$ cells with varying size between $19$ -- $100$~km, with $20$~km at the chromosphere. There are $504$ cells in each of the horizontal directions ($x,y$) with a constant grid spacing of $48$~km, resulting in a corresponding angular resolution of approximately $0.066$ arcsec (at a distance of 1 AU). 
The boundary conditions in the horizontal directions are periodic.
The simulation has the same setup as the publicly available version \citep{2016A&A...585A...4C}, but instead of the $10$~s cadence, a higher output cadence of $1$~s was used. 
The duration of the simulation is approximately 1~hour.
The publicly available version (at lower cadence) was for instance used to analyze the diagnostics at mm and submm wavelengths by \cite{2015A&A...575A..15L, 2017A&A...601A..43L} and to study the magnetic field topology by \cite{2017ApJS..229...11J}. The same high-cadence simulation that is used in the current work was previously used to study the diagnostic potential of solar ALMA observations by \cite{2020A&A...635A..71W}, the impact of spatial resolution on ALMA observations of the Sun by \cite{2021A&A...656A.68E} and the variation of the synthetic observables at mm-wavelengths and the magnetohydrodynamic properties of the atmosphere in connection to a propagating shock wave by \cite{2021RSPTA.37900185E}.

\subsection{Brightness temperature maps at mm-wavelengths corresponding to ALMA observations}\label{sec:method:rad_transfer}

In the current study we analyse the observables at wavelengths corresponding to ALMA receiver band~3 (centered at 3.0~mm / 100~GHz) and band~6 (centered at 1.3~mm / 239~GHz).
The observable continuum intensities at mm wavelengths were obtained by solving the equation for radiative transfer using the Advanced Radiative Transfer (ART) code \citep{art_2021} for each vertical column at each time step.
The calculations are performed at three wavelengths in each spectral receiver sub-band, at respective minimum, middle and maximum wavelengths, as indicated in Table~\ref{tab:frequencies}.
The intensities are then transformed to brightness temperatures using the Rayleigh-Jeans approximation \citep[see e.g.,][]{2013tra..book.....W}.
The resulting brightness temperature observables for each sub-band are then constructed by averaging the brightness temperatures at the three corresponding wavelengths.

\begin{table}[t!]
\caption{
Wavelengths and frequencies of the sub-bands of receiver bands 3 and 6 that are used for the radiative transfer calculations.
}
\label{tab:frequencies}
\begin{tabular}{lcccccc}
\hline
Sub-&\multicolumn{3}{c}{Wavelength [mm]}&\multicolumn{3}{c}{Frequency [GHz]}\\
band&min&mid &max& min & mid  & max \\
\hline                
SB3.1 & 3.1893 &3.2236& 3.2586 &  92.0 &93.0&  94.0 \\ 
SB3.2 & 3.1228 &3.1557& 3.1893 &  94.0 &95.0&  96.0 \\
SB3.3 & 2.8282 &2.8552& 2.8826 & 104.0 &105.0& 106.0 \\
SB3.4 & 2.7759 &2.8018& 2.8282 & 106.0 &107.0& 108.0 \\
 \hline
SB6.1 & 1.2978 &1.3034& 1.3091 & 229.0&230.0& 231.0 \\ 
SB6.2 & 1.2867 &1.2922& 1.2978 & 231.0&232.0& 233.0 \\
SB6.3 & 1.2137 &1.2187& 1.2236 & 245.0 &246.0& 247.0 \\
SB6.4 & 1.2040 &1.2088& 1.2137 & 247.0&248.0& 249.0 \\
\hline
\end{tabular}
\label{tab:almasubbands}
\end{table}

In the current work we look at the slope of the brightness temperature continuum within the receiver bands, which we derive by 
the \hea{difference of brightness temperature between the outermost receiver sub-bands},
SB4 and SB1, for each respective band (Table~\ref{tab:almasubbands}).
\hea{The} sub-band differences for bands~3 and 6 are calculated as
\begin{equation}\begin{split}
    \Delta T_\mathrm{b} |_\text{B3} &= T_\mathrm{b} |_\text{SB3.1} - T_\mathrm{b} |_\text{SB3.4},\\
    \Delta T_\mathrm{b} |_\text{B6} &= T_\mathrm{b} |_\text{SB6.1} - T_\mathrm{b} |_\text{SB6.4}     
\label{eq:SBdiffDef}
\end{split}\end{equation}
where $T_\mathrm{b} |_\text{SB3.1}$, $T_\mathrm{b} |_\text{SB3.4}$, $T_\mathrm{b} |_\text{SB6.1}$ and $T_\mathrm{b} |_\text{SB6.4}$ are the brightness temperatures for SB3.1, SB3.4, SB6.1 and SB6.4, respectively.


\begin{figure*}[t]
\sidecaption
\includegraphics[width=\textwidth]{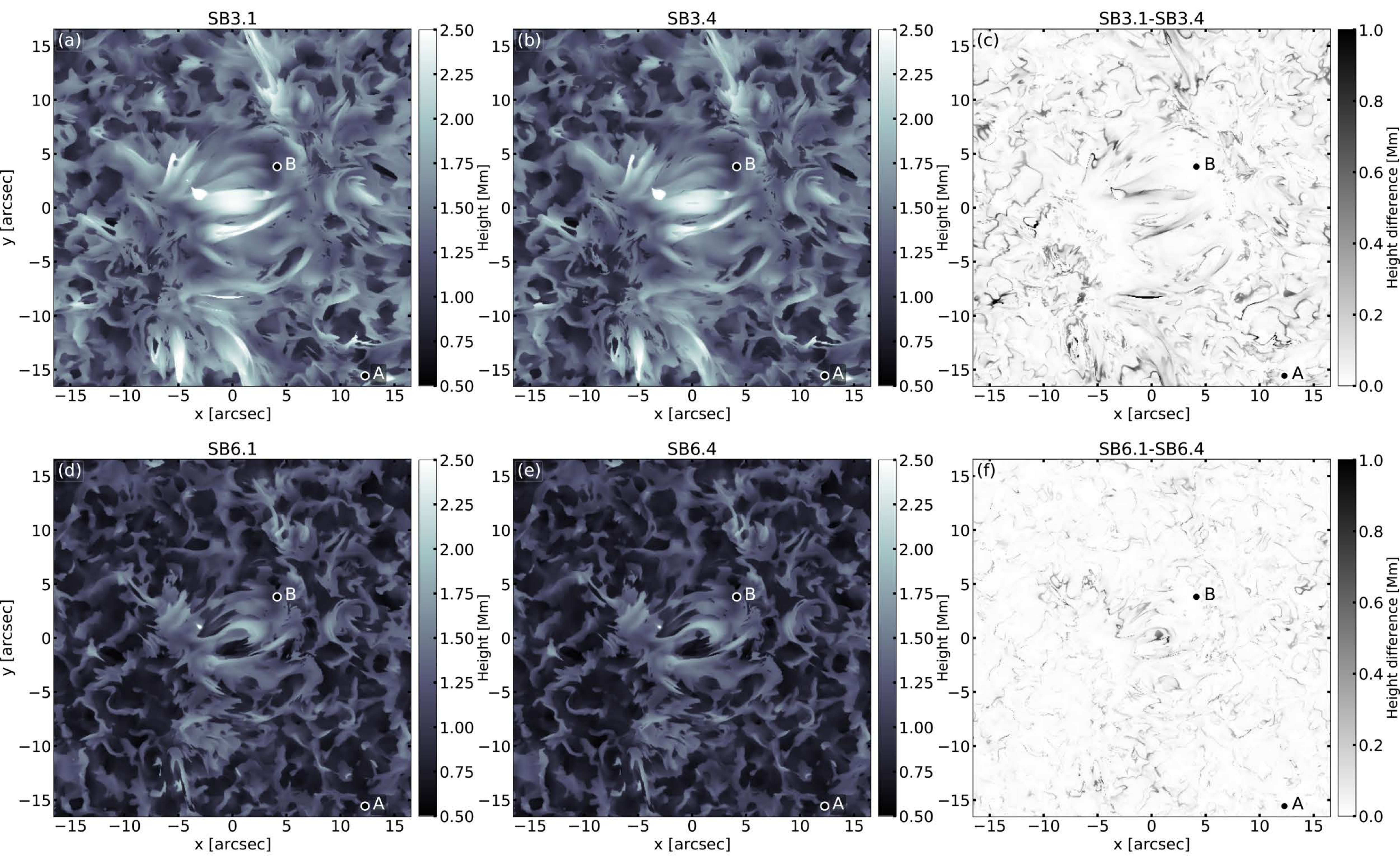}
\caption{\hea{The height of formation of the radiation at: (a) SB3.1 ($\tau_\text{3.2~mm}=1.0$), (b) SB3.4 ($\tau_\text{2.8~mm}=1.0$), (c) difference in height between SB3.1 and SB3.4, (d) SB6.1 ($\tau_\text{1.3~mm}=1.0$), (e) SB6.4 ($\tau_\text{1.2~mm}=1.0$) and (f) difference in height between SB3.1 and SB3.4. The selected locations, A and B, for time-series analysis (Figs.~\ref{fig:CFs_shockwave}--\ref{fig:CFs_network}) are marked for reference.
The color scale of the sub-band differences (c) and (f) are at all times positive (they are not absolute) and are capped to $1.0$~Mm from $1.42$~Mm at band~3 (c) and $1.17$~Mm at band~6 (f), in order to reveal the small scale structures.}
}
\label{fig:formation_heights_FOV}
\end{figure*}

\begin{figure*}[h]
\sidecaption
\includegraphics[width=\textwidth]{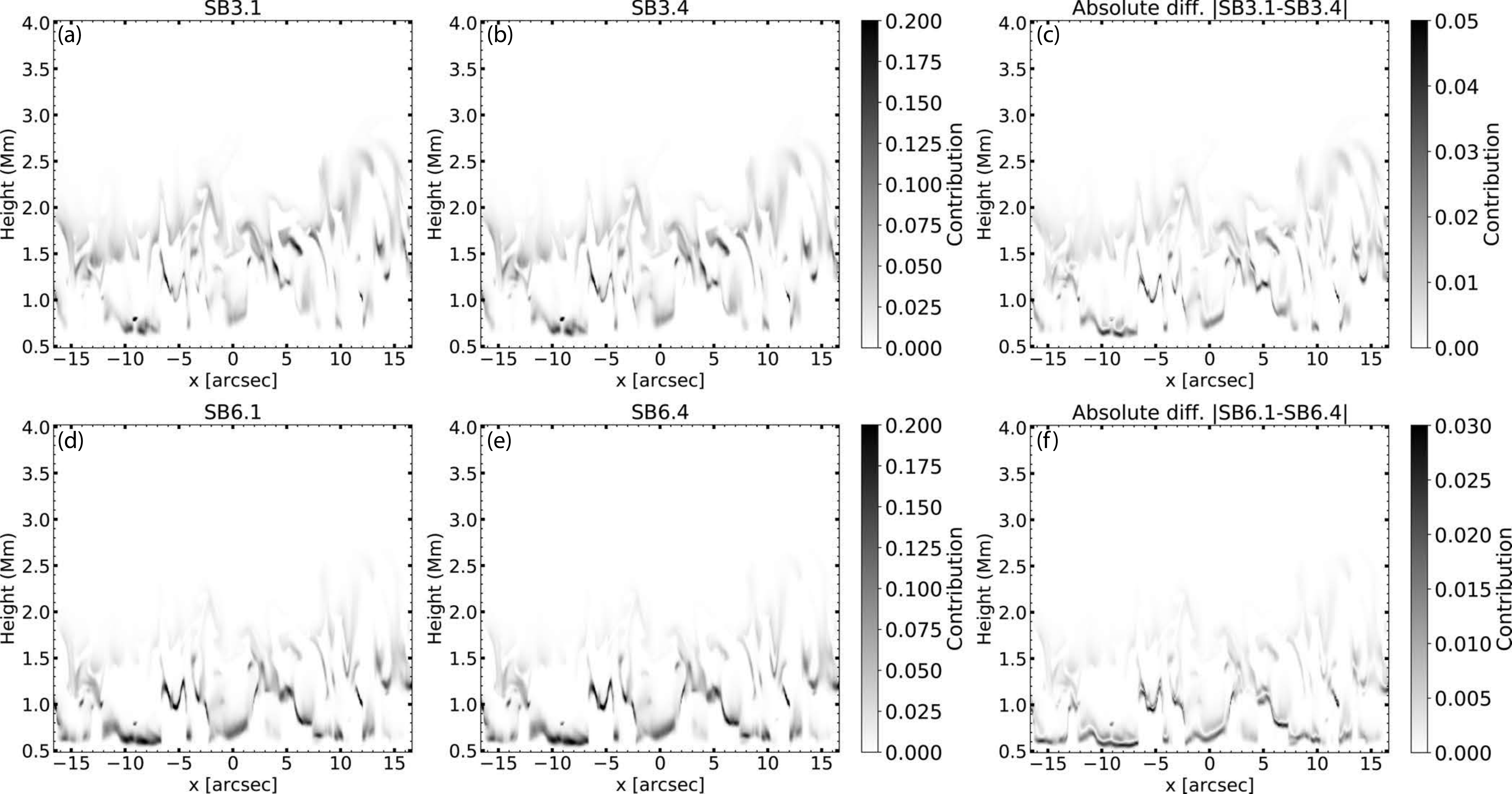}
\caption{Contribution \hea{to the total brightness temperature as function of height along a slit at} $y=-15$~arcsec of SB3.1 (a), SB3.4 (b), the absolute difference between the SB3.1-SB3.4 (c), SB6.1 (d), SB6.4 (e) and the absolute difference between the SB6.1-SB6.4 (f). Notice the different color scales at the different bands.
}
\label{fig:CF_SBdiff}
\end{figure*}


\subsection{\hab{Synthesizing mm-maps with the angular resolution of ALMA observations}}

The physical locations of the 12-m antennas in the interferometric array can be reconfigured and arranged in different configurations. 
This changes the distribution of the distances and angles between the antennas. \hea{The vector between the antennas is referred to as a baseline.} The distribution of the baselines determines at what scales and angles the interferometric array is measuring the target.
Longer baselines sample smaller spatial scales in the source and shorter baselines larger scales. 
The resulting angular resolutions of the so far offered antenna array configurations are on the same order or larger than the scales of typical small-scale dynamic events in the chromosphere, leaving much of the small-scale dynamics unresolved \citep{2020A&A...644A.152E}.
The more extended antenna configurations are therefore more favourable for resolving  chromospheric small-scale events \citep{2021A&A...656A.68E}.

\hab{
We degrade the brightness temperature maps (Sect.~\ref{sec:method:rad_transfer}) to the resolutions corresponding to interferometric ALMA observations.
Synthesised beams with realistic sizes and shapes are acquired by performing synthetic interferometric ALMA observations, using the \textit{simobserve} tool included in the \textit{Common Astronomy Software Applications package} \citep[CASA; v.6.1.0;][]{2007_McMullin_CASA},
with different antenna array configurations and frequencies of the receiver sub-bands (Table~\ref{tab:almasubbands}).
These beams are then used to degrade the brightness temperature maps by convolution.
See also \cite{2021A&A...656A.68E, 2020A&A...644A.152E} where the same process was used, for more details.
Because of the frequency dependency of the resolution, the receiver sub-bands have slightly different resolution.
The resulting resolutions for each combination of array configurations C1--C7 and sub-bands 1 and 4 of receiver bands 3 and 6 are for reference listed in Table~\ref{tab:appendix_clean_beams}.}

\subsection{Observational data}
An observational data set taken with ALMA in band~$3$ \hea{(project ID: 2017.1.00653.S)} in a quiet Sun region close to the disc centre is used as comparison and validation of the results seen in the simulations. 
The measurements were performed in the so far most commonly \hea{used
antenna configuration for solar observations at band~3 (C3), resulting in 
a clean beam (main lobe of the point spread function; PSF) with} an average size of $1.8\arcsec\times2.6\arcsec$ along the minor and major axis, respectively. 
The measurements were made \hea{on April 12, 2018 between 15:52~UT and 16:24 UT} in four blocks of about $10$~min each, separated by calibration gaps of about $2$~min. 
The data was reduced with the Solar ALMA Pipeline (SoAP; Szydlarski et al., in prep.) that largely automatizes the process of phase self-calibration\ha{,} deconvolving process using the CLEAN algorithm \citep{1974A&AS...15..417H}. \ha{The interferometric data was also combined with the total power measurements \citep{2017SoPh..292...88W, 2017SoPh..292...87S} to acquire absolute temperatures. See}  \cite{2020A&A...635A..71W, 2020A&A...644A.152E, 2021arXiv210902374H} for summarized descriptions of the employed techniques of the image reduction of the high cadence time series.
\ha{Images} are constructed for each sub-band (Table~\ref{tab:almasubbands}) with an integration of $1$~s. \hea{The same dataset with images reconstructed from the full receiver band is available in the Solar ALMA Science Archive \citep[SALSA; ][]{2021arXiv210902374H}.}


\section{Results}\label{sec:results}

\subsection{Formation heights of radiation and corresponding plasma temperature}
\label{sec:formation_heights}

\hea{The formation heights of the radiation of the sub-bands of bands~3 and 6} 
are \hea{often} determined as the heights where the optical depth is unity for the corresponding wavelength, i.e. 
$\tau_{3.2~\text{mm}}=1.0$, $\tau_{2.8~\text{mm}}=1.0$, $\tau_{1.3~\text{mm}}=1.0$ and $\tau_{1.2~\text{mm}}=1.0$, respectively. 
As can be seen from the resulting formation heights across the entire field-of-view \ha{of the simulation} at the time of $t=1600$~s in Fig.~\ref{fig:formation_heights_FOV},     
the average formation height of the  millimeter continuum is wavelength dependent, where the height of formation is larger for longer wavelength \citep[see e.g.,][]{1981ApJS...45..635V,2004A&A...419..747L,2007A&A...471..977W,2013tra..book.....W,2016SSRv..200....1W}.
%
Statistics of the formation heights of the synthetic radiation at the ALMA receiver bands from the same \textit{Bifrost} 3D model that is used here (Sect.~\ref{sec:method-bifrost_model}) is presented by \cite{2021A&A...656A.68E}. Their study considers the average values over the whole receiver bands. They confirm that the height of formation increases on average with wavelength and that it is on average larger at the network patches than at the surrounding quiet Sun. 

The corresponding differences in height of formation of the sub-bands of bands 3 and 6 (SB3.1-SB3.4 and SB6.1-SB6.4) are given in Fig.~\ref{fig:formation_heights_FOV}c and f, respectively.
\hea{The difference of formation height (i.e. $z(\tau=1.0)_\lambda$ at the respective wavelength)
between the sub-bands is at all times positive with SB3.1 forming higher up than SB3.4 and SB6.1 forming higher up than SB6.4.}
This is the case at all times and locations throughout the entire simulation box (\hea{see Sect.~\ref{sec:method-bifrost_model}}). 
We want to point out that 
this trend is also seen at all times for each of the ALMA receiver bands between band~3 (100~GHz) -- band~10 (867~GHz) \citep[see][for details]{2021A&A...656A.68E}. 
Please note that data for these bands has been calculated, too, 
but for simplicity, we do not include all bands in the detailed analysis here.
\hea{The consistent order of formation height of the sub-bands} naturally comes from the wavelength dependency of the opacity.
We thus have for the formation height $z$,
\begin{equation}
    z(\tau_{\text{SB}1}=1.0) > z(\tau_{\text{SB}4}=1.0)
\end{equation}
where the radiation of $\text{SB}1$ has longer wavelength than that of $\text{SB}4$ for each band (Table~\ref{tab:almasubbands})

An optical depth of $\tau(\nu)=1.0$ can serve as a good first approximation of the height of formation of the radiation,
\hea{for instance in the case of 1D atmospheric models, such as the static VAL model \citep{1981ApJS...45..635V} and also with} introduction of plane-parallel propagating shocks \citep[see e.g.,][and references therein]{2004A&A...419..747L, 2002ApJ...572..626C}.
However, in a 3D environment with more complex dynamics due to propagating and interfering waves, 
there can be contribution to the intensity of the ALMA bands from  several major components at different heights along the line of sight.
Examples are shown in Fig.~\ref{fig:CF_SBdiff}, where the contribution functions of SB3.1, SB3.4, SB6.1 and SB6.4, respectively,  are shown for a slit across the FOV at $y=-15$~arcsec.

The \hea{change of opacity from one sub-band to another is not very large and the} major components of the contribution functions of the two sub-bands originate largely from the same heights, \hea{at a given location},
which is the case for both receiver bands~3 and 6 \hea{((Fig.~\ref{fig:CF_SBdiff})}.
It is thus these heights, displayed in (Fig.~\ref{fig:CF_SBdiff}c and f), that are probed by the sub-bands and for which the brightness temperature differences (slope of the brightness temperature continuum) provide  information on the slope of the local plasma temperature.  
We therefore, for simplicity, focus on the contribution function of a single sub-band in each receiver band in the following in-depth analysis.  
\hea{Worth noting is that the components of the contributions to the intensities of bands~3 and 6 show large and overlapping height ranges (Fig.~\ref{fig:CF_SBdiff})}
Band~3 more often shows a strong component at large altitudes simultaneous with a component at lower altitudes, while band~6 shows a more concentrated contribution to the lower component and only a weak upper component.

\begin{figure}[!t]
\includegraphics[width=\columnwidth]{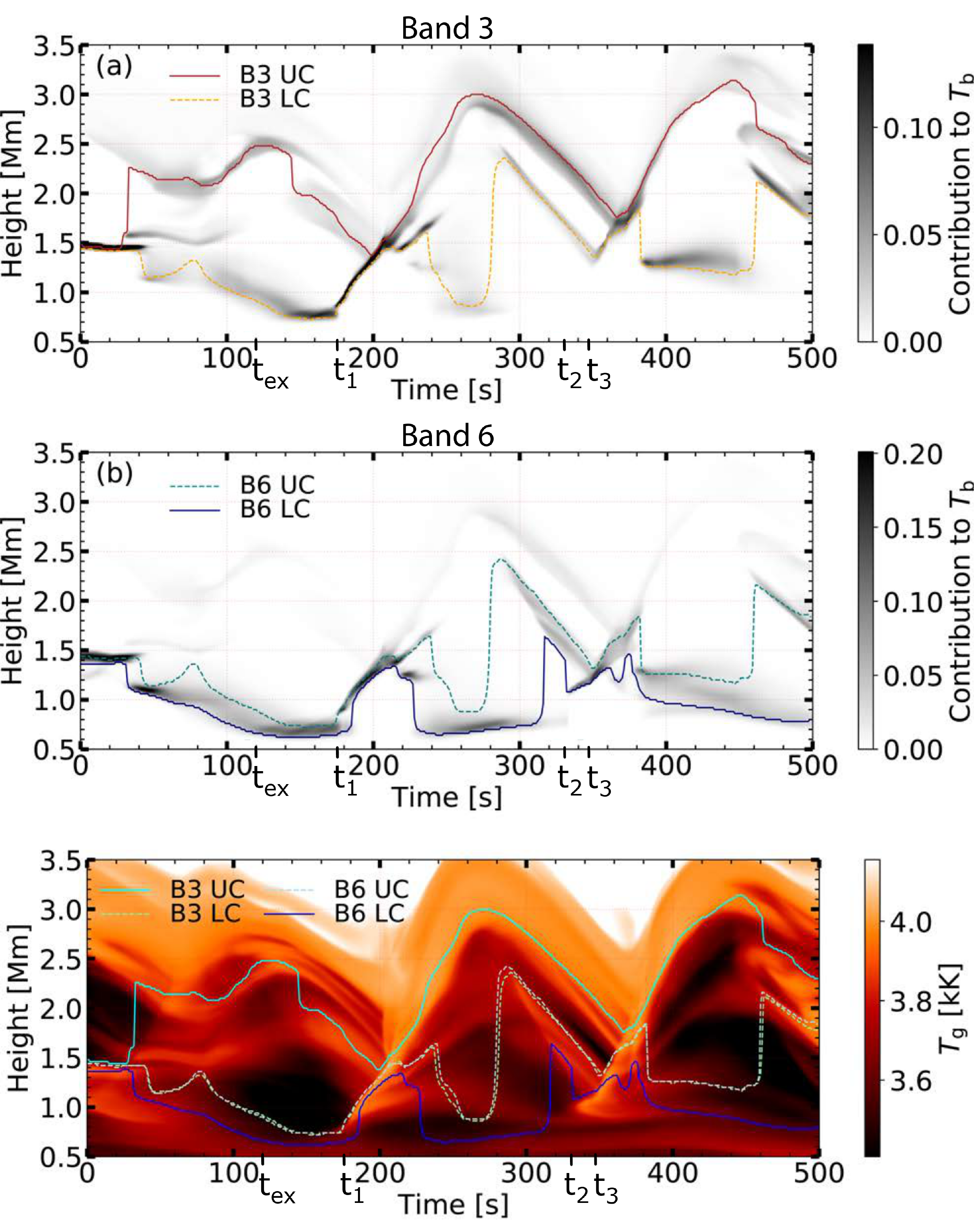}
\caption{Time evolution of contribution functions and plasma temperature for a column in the quiet Sun, \hea{at location A (Fig.~\ref{fig:formation_heights_FOV}), with the shock wave event studied in detail in \cite{2021RSPTA.37900185E}}. 
\hea{Contribution function of band~3 (a) and band~6 (b) with the heights of $0.25$ and $0.80$ times the contribution function integrated from the top marked by the dashed and solid lines, respectively.} 
(c) Plasma temperature with \hea{each} of the representative heights of the two components of bands~3 and 6 marked for reference.
\ha{The ticks t$_{\mathrm{ex}}$ and t$_1$ -- t$_3$ mark the times for in-depth discussion in connection to Fig.~\ref{fig:tg_CFs_t120} and Fig.~\ref{fig:Tb_ev_ex}}.
}
\label{fig:CFs_shockwave}
\end{figure}

\begin{figure}[!t]
\includegraphics[width=0.9\columnwidth]{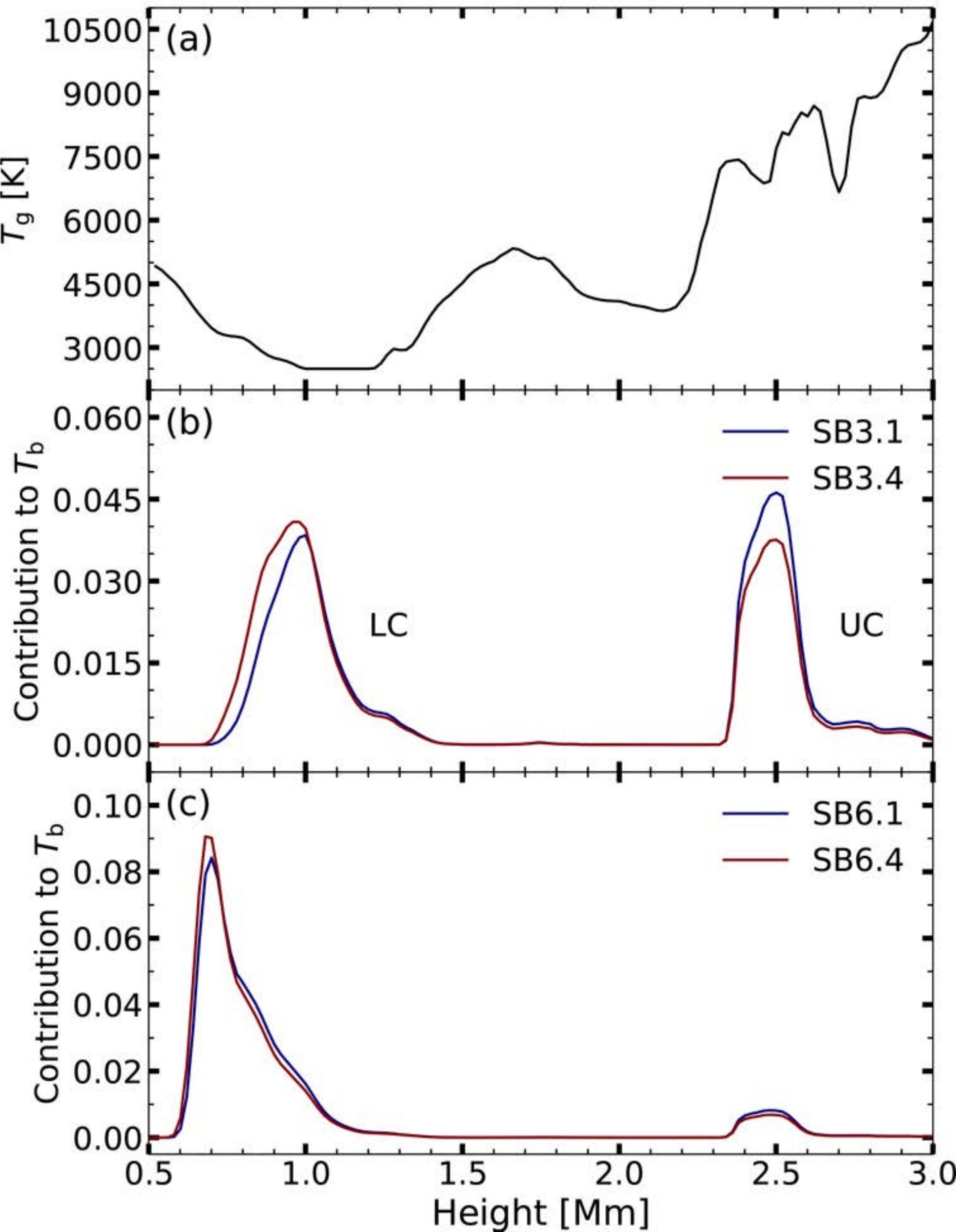}
\centering
\caption{Plasma temperature and contribution functions plotted against height in the atmosphere for one time step, \ha{$t_\mathrm{ex}=120$~s, of the shock wave (cf. Fig.~\ref{fig:CFs_shockwave})} \hea{at location A (Fig.~\ref{fig:formation_heights_FOV})}. 
(a) Plasma temperature as function of height. (b) Contribution functions of SB3.1 and SB3.4, each showing a lower and an upper component, marked LC and UC, respectively.
(c) Contribution functions of SB6.1 and SB6.4. 
}
\label{fig:tg_CFs_t120}
\end{figure}


In Fig.~\ref{fig:CFs_shockwave}, the evolution of the contribution functions of SB3.1, SB3.4, SB6.1 and SB6.4, respectively are shown for a time-series ($t=0$--$500$~s) \hea{of propagating shock waves in quiet Sun conditions \hea{\citep[see][for identification of regions in the simulaiton]{2021A&A...656A.68E}}, \ha{in the corner of the simulation box} at $(x,y)=(-15.6,12.3)$~arcsec (location A in Fig.~\ref{fig:formation_heights_FOV}).}
The beginning of the time-series ($t=0-300$~s) is presented in detail by \citet{2021RSPTA.37900185E} including the excitation of a shock wave around $t=175$~s. 
Here we show an additional $200$~s of  the time series at the same location, which includes the excitation of an additional shock around t=$350$~s.
\hea{The shock shows a magnitude of brightening excess that is commonly seen for events in the simulation under quiet Sun conditions \citep{2021A&A...656A.68E}.}
In the time series at this location,  band~3 shows two major components in the contribution to the brightness temperatures at most times (Fig.~\ref{fig:CFs_shockwave}a). 
We refer to the two components as the upper comonent (UC) and the lower component (LC). The heights of the two components are \hea{at most times well estimated} by $0.25$ (UC) and $0.80$ (LC) of the integrated contribution function from the top \hea{(marked by dashed and solid lines in Fig.~\ref{fig:CFs_shockwave})}.
There are a couple of exceptions where the contribution is fairly concentrated in a single component, which is when the upwardly propagating shock front intersects with the down-falling gas at $t=200$~s and $t=370$~s. 
The vertical velocities of the time-series are shown in Appendix~\ref{appendix:vertical_vel}  for reference.
At these times, \hea{the UC and LC}
\hea{of band~3 merge and sample the same height range and, as we will see below (Sect.~\ref{sec:Tb-time-series_example})}, the brightness temperature is large at these points.

\hea{While band~3 displays significant contribution from both the LC and the UC at significantly different heights at most times (Fig.~\ref{fig:CFs_shockwave}a), 
band~6 shows a mix between occasions with
weaker contributions from the upper layers, leaving both the LC and UC sampling the concentrated contribution function at similar heights (i.e., $t=120$~s in Fig.~\ref{fig:CFs_shockwave}b) and with stronger contributions from the upper layers, giving rise to a significant difference in height between the LC and UC (i.e., $t=300$~s).
The relative contribution between the LC and the UC is complex and dependent on the wavelength of the radiation and the small-scale dynamic structures in the atmosphere.}
\ha{In the quiet Sun conditions, the propagating shocks} give rise to the conditions where there is an UC of the contribution to the brightness temperatures 
(up to $\sim$3.0--3.5~Mm at band~3 and up to $\sim$2.5--3.0~Mm at band~6 of the selected example in Fig.~\ref{fig:CFs_shockwave}).
\hea{The} UC is seen both when the mm wavelength radiation track the propagating shock upwards and in the corresponding post-shock regions.
\ha{However, the shocks does not give rise to a dense enough upper layer to be optically thick in neither of bands~3 or 6, resulting in that a lower layer is simultaneously sampled, hence the presence of the LC. 
The cool low density post-shock regions can separate the LC from the UC by a couple Mm. See \cite{2021RSPTA.37900185E} for further details on the atmospheric perturbation by the shock.}

\hea{The shock starting around $t=178$~s is sampled by both band~3 (LC) and band~6 simultaneously at a  height $z\approx 0.7$~Mm (Fig.~\ref{fig:CFs_shockwave}), 
and \hea{is tracked} up to around $1.4$~Mm at $t=210$~s.
Thereafter, the UC of band~6 and LC of band~3 track a secondary component of the shock \citep[hook-shaped feature in Fig.~\ref{fig:CFs_shockwave}), see also][]{2021RSPTA.37900185E} up to about $1.7$~Mm at $t=240$~s, and the UC of Band~3 track the shock front all way up to about $3.0$~Mm. 
}

As an illustrative example, the contribution to the intensities as function of height for bands~3 and 6 at $t=120$~s of the time-series is displayed together with the plasma temperature as function of height in Fig.~\ref{fig:tg_CFs_t120}, on the same $x$-axis for ease to read out the slope of the plasma temperature at the sampled heights. 
Here, the contribution functions of SB6.1 and SB6.4 each \hea{show one major component, centered around $0.7$~Mm, and only a very small contribution at heights around $2.5$~Mm.}
\hea{At the heights around $0.7$~Mm sampled by band~6}, the plasma temperature decreases with increasing height (Fig.~\ref{fig:tg_CFs_t120}a).
\hea{The contribution functions of} SB3.1 and SB3.4 at $t=120$~s, show on the other hand each two major components, the LC around $1.0$~Mm and the UC around $2.5$~Mm.
\hea{Around the heights of the LC, there is a very slight decrease of plasma temperature with increasing height and around of the heights of the UC the plasma temperature increases (on average over the UC) with height  (Fig.~\ref{fig:tg_CFs_t120}a).}
\hea{The slope of the gas temperature at the sampled layers is reflected by the brightness temperature difference between the sub-bands, which we discuss further in Sect.~\ref{sec:Tb-time-series_example}.}

\begin{figure}[t]
\includegraphics[width=\columnwidth]{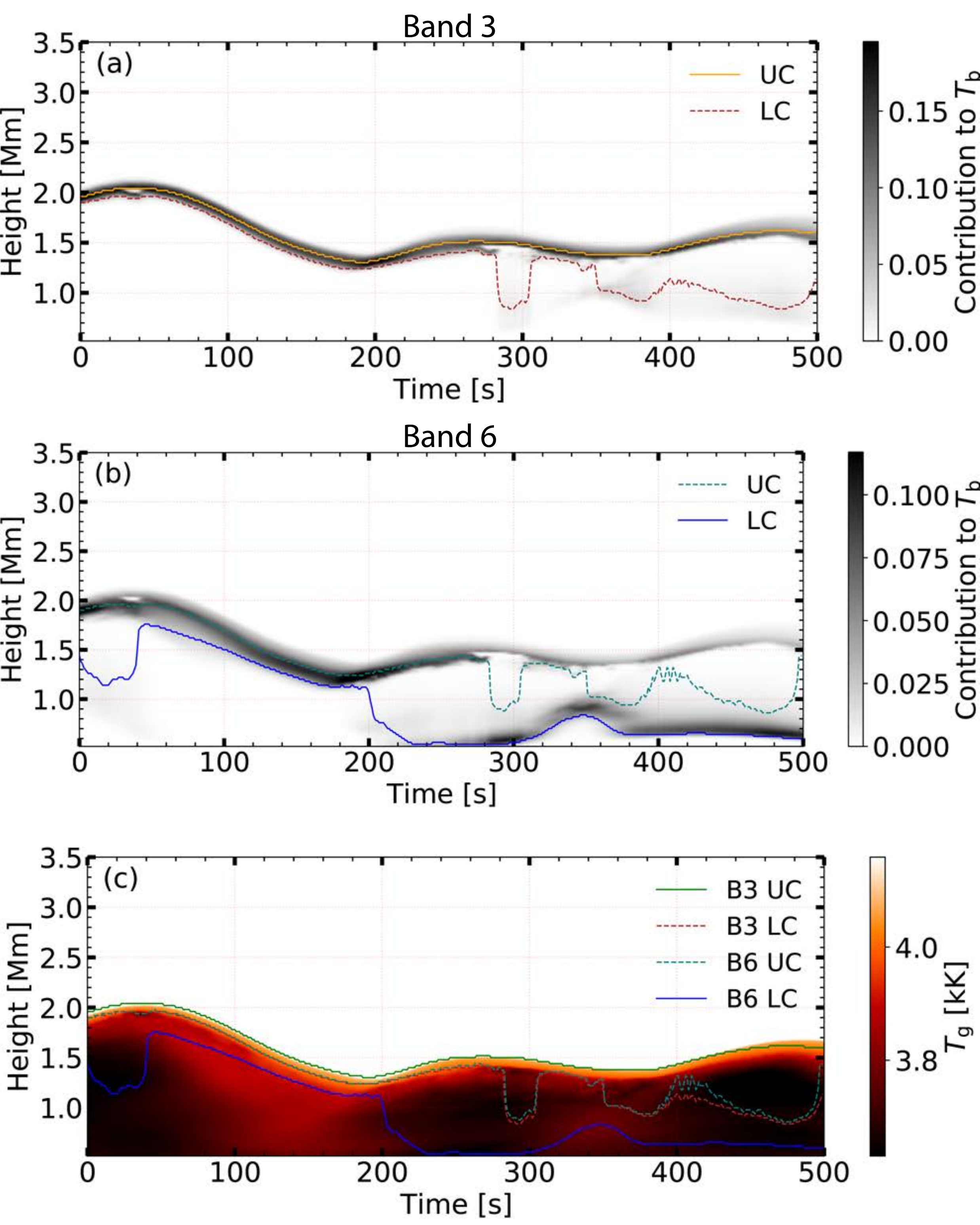}
\caption{Time evolution of contribution functions and plasma temperature for a column in the network patches at (x,y)=(3.8,4.2)~arcsec, marked as location B at the FOV maps (Fig.~\ref{fig:formation_heights_FOV}). Top panel: Contribution function of band~3 with the heights of 0.25 and 0.8 times the contribution function integrated from the top marked by the dashed and solid lines, respectively. 
Middle panel: Same as top panel for band~6. 
Bottom panel: Plasma temperature with representative heights of formation of bands~3 and 6 marked with dashed and solid lines, respectively. The formation height at $\tau=1.0$ of both sub-bands are indicated in respective row for ease of comparison.}
\label{fig:CFs_network}
\end{figure}

Similar to the example of shocks under quiet Sun \hea{conditions (Fig.~\ref{fig:CFs_shockwave})}, a time-series of the contribution functions and the corresponding plasma temperature at a location in the network region (location B in Fig.~\ref{fig:formation_heights_FOV}) is given in Fig.~\ref{fig:CFs_network}. 
\hea{During the first 200~s, both bands~3 and 6 show a contribution functions with one well-concentrated major component at a similar height.
}
\hea{Band~6 however show} two major components to the contribution function towards the end of the time-series. The formation heights at location~B in the network region is less dynamic than at location~A in the quiet Sun. The corresponding vertical velocities \hea{for the time-series} are given in Appendix~\ref{appendix:vertical_vel}.

\subsection{Synthetic observables, difference of brightness temperature between sub-bands}\label{sec:brightness temp.}

\subsubsection{Variation over the field of view}\label{sec:Tb-FOV}

\begin{figure*}[tbh]
\sidecaption
\includegraphics[width=\textwidth]{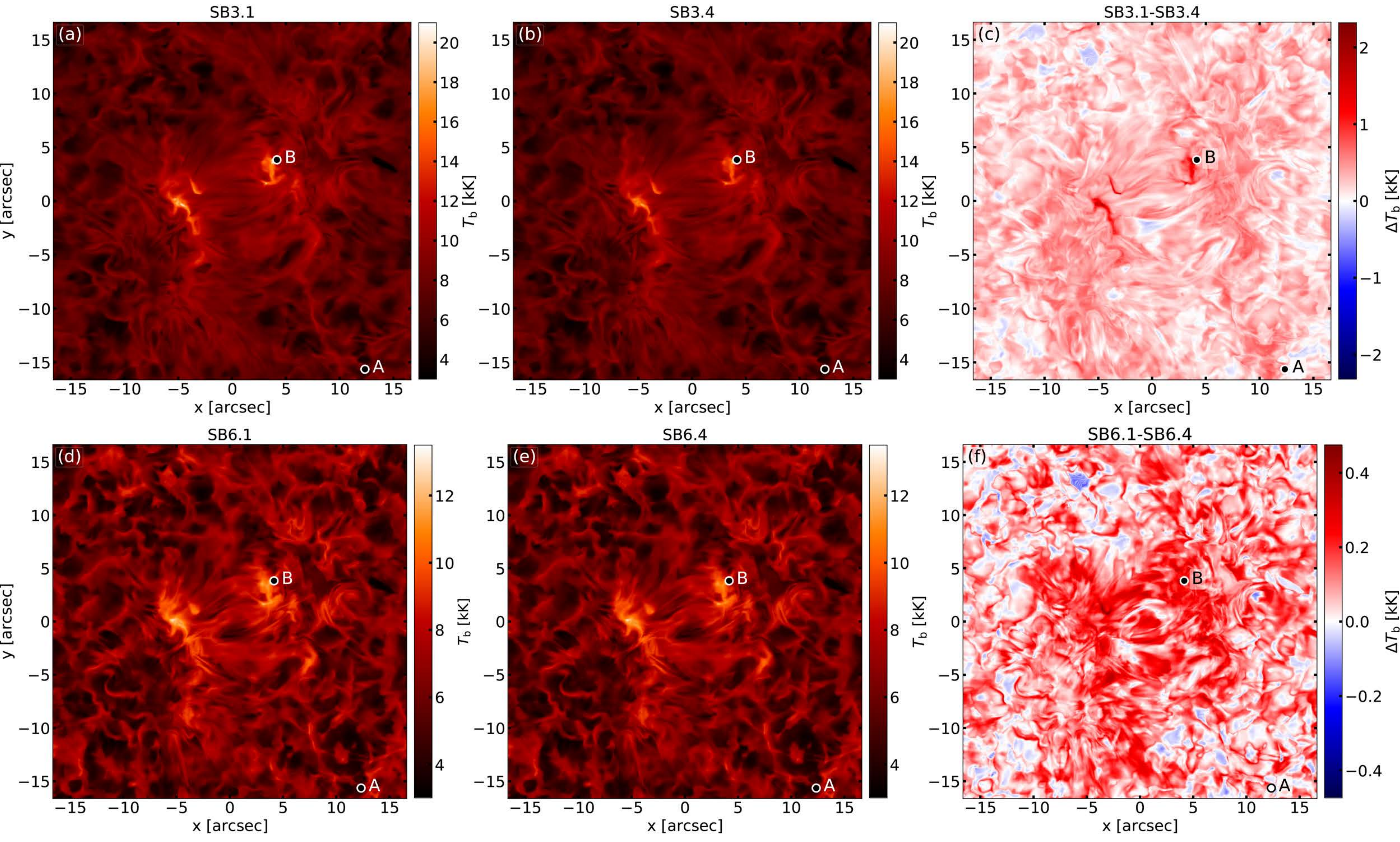}
\caption{Brightness temperatures of sub-bands of bands~3 and 6. (a) $T_\mathrm{b} |_\text{SB3.1}$, (b) $T_\mathrm{b} |_\text{SB3.4}$, (c) $\Delta T_\mathrm{b} |_\text{B3}$ ($T_\mathrm{b} |_\text{SB3.1} - T_\mathrm{b} |_\text{SB3.4}$),
(d) $T_\mathrm{b} |_\text{SB6.1}$, (e) $T_\mathrm{b} |_\text{SB6.4}$, (f) $\Delta T_\mathrm{b} |_\text{B6}$ ($T_\mathrm{b} |_\text{SB6.1} - T_\mathrm{b} |_\text{SB6.4}$). 
The selected locations, A and B, for time-series analysis are marked for reference. Please note the different color scales of the two receiver bands.
}
\label{fig:Tb_FOV}
\end{figure*}

The synthetic observables of the sub-bands of bands~3 and 6  \hea{at $t=1600$~s} are presented in Fig.~\ref{fig:Tb_FOV}.
The brightness temperatures ranges are in line with those for the corresponding full-band maps, which are analysed in detail for different areas in the field of view by \cite{2021A&A...656A.68E}.
For band~3 the average brightness temperatures over the entire FOV are 7710~K at SB3.1 and only 7430~K at SB3.4, with values of up to 21050~K at a few locations in the central network patches.
\ha{These values \citep[see][for corresponding full-band temperature distributions]{2021A&A...656A.68E} agree well with average values reported for observational full-band~3 ALMA data of the quiet Sun, e.g., 
7056~K for internetwork (IN) and 7694~K for network (NW) regions by \cite{2020A&A...640A..57A} and 7228~K (IN) and 7588~K (NW) by \cite{2020A&A...635A..71W}. They are also in line with observations with other instruments, such as the Nobeyama 45~m telescope, measuring $7700\pm 310$~K at 2.6~mm (close to SB3.4; Table~\ref{tab:almasubbands}) averaged over the whole solar disk \citep{2017SoPh..292...22I}}.
Band~6 shows significantly lower average temperatures of 6190~K at SB6.1 and 6090~K at SB6.4, and maximum temperatures of only up to 13310~K.

The difference of the brightness temperatures between SB3.1 and SB3.4, $\Delta T_\text{b}|_{\text{B}3}$ (see Eq.~\eqref{eq:SBdiffDef}), across the FOV of Fig.~\ref{fig:Tb_FOV}c shows values between $-377$~K and -$2320$~K. 
\hea{There} are distinct differences in the magnitude and the sign of $T_\mathrm{b}$ sub-band differences with different magnetic field topology. 
That effect is expected as the magnetic field interferes with propagating shock fronts resulting in reflection or transformation into other magneto-acoustic wave modes \hea{and canopies of magnetic fibrils can obscure the view of the shocks at lower heighs}. As a result, shock front signatures at millimeter wavelengths are not seen as frequently at areas with larger magnetic field strength \citep{2020A&A...644A.152E}, such as the magnetic field footpoints in the central FOV (Fig.~\ref{fig:Tb_FOV}).

The largest $\Delta T_\text{b}|_{\text{B}3}$ are seen at the few locations displaying the highest temperatures in the central network patches. In the surrounding region of network patches, a substantial positive $\Delta T_\text{b}|_{\text{B}3}$ of several hundreds K is displayed at most locations. 
\hea{There} are also occasionally occurrences with negative sub-band $T_\mathrm{b}$ difference, most notable in band 3, in the area between the magnetic field footpoints where the magnetic loops are mostly horizontally inclined at the heights of ALMA bands~3 and 6 (\cite{2017ApJS..229...11J}; see also \hea{Fig.}~\ref{fig:formation_heights_FOV}).
Most of the occasions with low or even negative $\Delta T_\text{b}|_{\text{B}3}$ values are seen in the corners of the FOV, which are representative of quiet Sun conditions.  
The sub-band differences in band~6, $\Delta T_\text{b}|_{\text{B}6}$, show a similar picture but with lesser magnitudes, with values only between $-182$~K and $474$~K.  

As a result of that the order of the formation height of the sub-bands consistently remains the same (Sect.~\ref{sec:formation_heights}), the slope of the continuum brightness temperature consequently reflects the slope of plasma temperature at the sampled heights. 
A positive $\Delta T_\mathrm{b}$ indicates a plasma temperature increasing with height, at the sampled layer. A positive value is thus in agreement with the classical models of the chromosphere, where a stratification of average temperature is increasing with height \citep[see e.g.,][]{1981ApJS...45..635V}.
A negative $\Delta T_\mathrm{b}$ indicates a decreasing plasma temperature with increasing height, at the sampled \hea{layer}.

In the network patches in the central parts of the FOV (Fig.~\ref{fig:Tb_FOV}c), which show large $\Delta T_\text{b}|_{\text{B}3}$ values, the differences in height of formation of the sub-bands ($z(\tau_{\text{SB}3.1}=1.0) - z(\tau_{\text{SB}3.4}=1.0)$) are low (Fig.~\ref{fig:formation_heights_FOV}c), indicating steep gradient of the local plasma temperature increasing with height at the sampled layers of band~3.

In the quiet Sun conditions in the outskirts of the FOV \hea{of the synthetic maps}, there is a mix of positive and negative $\Delta T_\text{b}$ values at both bands~3 and 6.
The gradient of the brightness temperature continuum and therefore also the occurrence of \hea{negative continuum slopes} are coupled to the dynamics in the atmosphere and the \hea{negative continuum slopes} are mostly occurring at locations towards the corners of the FOV (Fig.~\ref{fig:Tb_FOV}), showing Quiet Sun conditions \hea{where} the dynamics is largely affected by propagation of shocks. 
The \hea{negative continuum slopes} at band~3 ($\Delta T_\mathrm{b} |_\text{B3}$) is stronger than in band~6 ($\Delta T_\mathrm{b} |_\text{B6}$), but band~6 shows more occurrences and more small-scale structures of the \hea{negative continuum slopes} than band~3.

\subsubsection{Time-series of brightness temperatures at selected locations}\label{sec:Tb-time-series_example}

\paragraph{Quiet Sun with propagating shock waves at location A.}

\begin{figure*}[h]
\includegraphics[width=\textwidth]{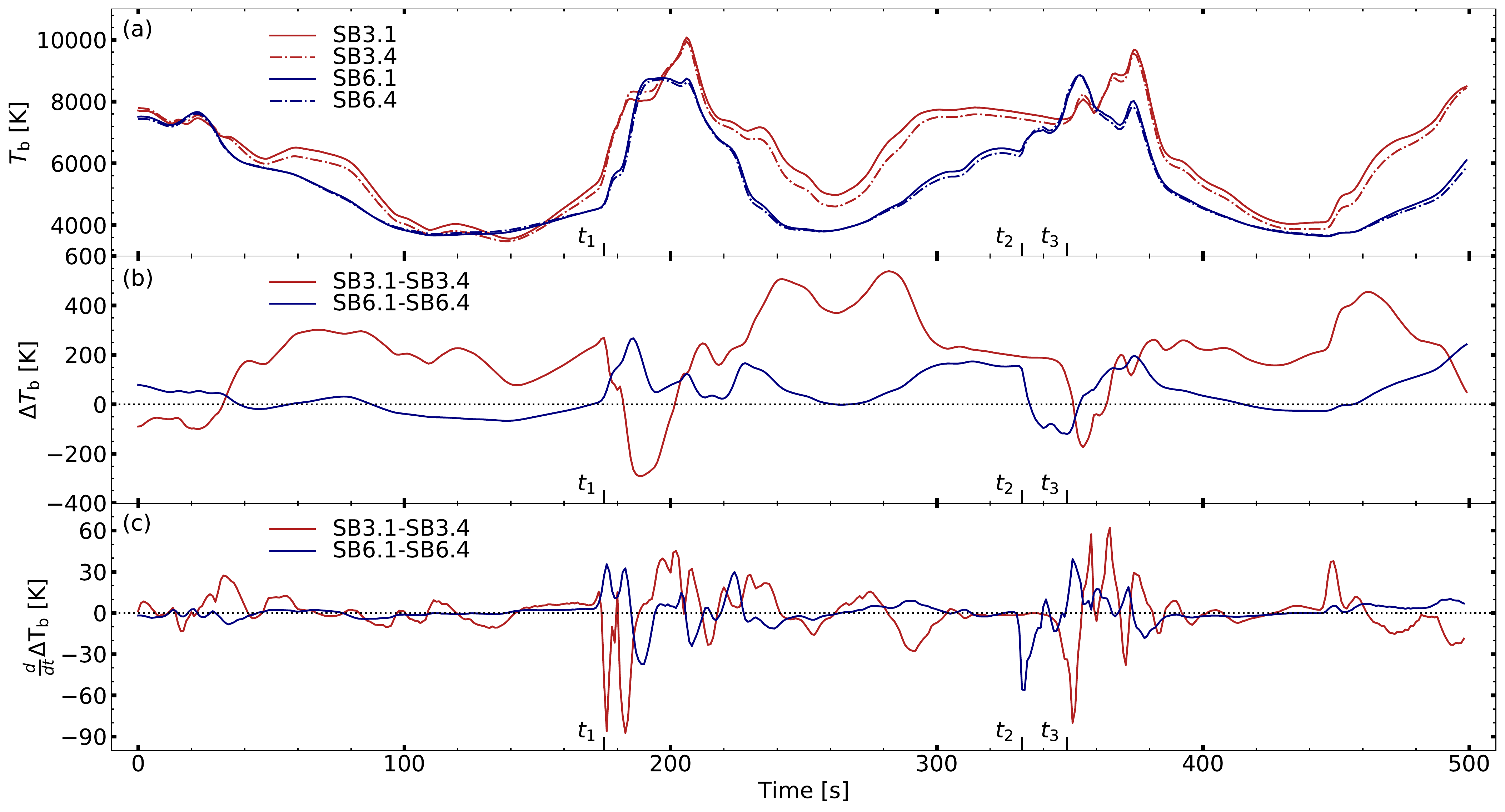}
\caption{Evolution of brightness temperatures and sub-band differences at the location of the propagating shock wave (location A) between $t=0$--$500$~s.
(a) The brightness temperatures of SB3.1, SB3.4, SB6.1 and SB6.4.
(b) Brightness temperature differences between the sub-bands  $\Delta T_\mathrm{b} |_\text{B3}$ ($T_\mathrm{b} |_\text{SB3.1} - T_\mathrm{b} |_\text{SB3.4}$) and $\Delta T_\mathrm{b} |_\text{B6}$ ($T_\mathrm{b} |_\text{SB6.1} - T_\mathrm{b} |_\text{SB6.4}$).
(c) Time-derivative of the sub-band differences $\Delta T_\mathrm{b} |_\text{B3}$ and $\Delta T_\mathrm{b} |_\text{B6}$.
In each panel, the time where the different bands starts to sample the shock waves are marked with the tick marks $t_1$ (bands~3 and 6), $t_2$ (band~6) and $t_3$ (band~3).}
\label{fig:Tb_ev_ex}
\end{figure*}

The temporal evolution of the brightness temperatures of the sub-bands of bands~3 and 6, at the location of propagating shock waves in the quiet Sun (location A in Fig.~\ref{fig:Tb_FOV}) is shown in Fig~\ref{fig:Tb_ev_ex}a between $t=0$--$500$~s.
Propagating shock fronts give rise to an increase in temperature and therefore brightness temperature at millimeter wavelengths \citep{1992ApJ...397L..59C, 2004A&A...419..747L, 2006A&A...456..713L, 2021RSPTA.37900185E, 2007A&A...471..977W}.
The beginning of the brightness temperature increase of the first shock front happens almost simultaneously for both bands~3 and 6 around $t_1=178$~s \hea{while there is a delay from sampling the second shock} front at $t_2=332$~s \hea{(around $z=1.1$~Mm) at} band~6 \hea{to} $t_3=345$~s \hea{(around $z=1.27$~Mm) at} band~3.
\hea{This accounts for a vertical shock propagation speed of about $13.1$~kms$^{-1}$, much in agreement with the speed of the first shock \ha{around $t_1=178$~s} \citep{2021RSPTA.37900185E}.
The delay between the sub-bands is however small $\sim1-2$~s at both bands, leaving the signatures barely resolved by the vertical grid spacing of the simulation (Sect.~\ref{sec:method-bifrost_model}).}
Both bands~3 and 6 largely samples at low heights ($\sim0.7$~Mm) right before they start to probe the first shock around $t_1=178$~s (Fig.~\ref{fig:CFs_shockwave}). 
Consequently, there is no delay of the signatures in the observables between the bands (Fig.~\ref{fig:Tb_ev_ex}). 
Although before the second shock front (e.g., around $t=300~s$), band~3 only samples the down-falling plasma while band~6 samples both the down-falling plasma and also partly plasma at low heights around $0.7$~Mm and therefore captures the shock front about $13$~s earlier than band~3.
This delay of $13$~s does not directly reflect the propagation speed of the shock front, but it instead gives an indication on the difference in opacity of the upper component measuring the down-falling plasma between bands~3 and 6.

The \hea{slope of the continuum}
is sensitive to sudden changes in the local \hea{atmosphere. 
Propagating} shock waves therefore show themselves in the time-evolution of the sub-band differences (Fig.~\ref{fig:Tb_ev_ex}b).
The brightness temperature differences between the sub-bands illustrated in Fig.~\ref{fig:Tb_ev_ex}b reflect the slope of the plasma temperatures, at the heights sampled by each receiver band, shown in Fig.~\ref{fig:CFs_shockwave}c.
At band~3, there is a sub-band \hea{difference} of $\mathrm{SB}3.1-\mathrm{SB}3.4$, with negative $\Delta T_\mathrm{b} |_\text{B3}$ values down to $-291$~K, between approximately $t=180$--$200$~s, while the first shock front is probed.
Band~6 on the other hand, shows an increase in sub-band difference.
Prior to the first shock front at $t_1=178$~s, band~6 is probing the plasma around 0.7~Mm (Fig.~\ref{fig:CFs_shockwave}), showing a negative sub-band difference down to about $-70$~K (Fig.~\ref{fig:Tb_ev_ex}b), indicating a decreasing temperature with increasing height at the sampled layers, which is seen to be the case in Fig.~\ref{fig:CFs_shockwave} and in Fig.~\ref{fig:tg_CFs_t120} of the specific example \ha{at $t_\mathrm{ex}=120$~s.
Band~3} is probing the down-falling plasma prior to the shock front at $t_1=178$~s, from about 2.5~Mm to 1.5~Mm and shows a large positive sub-band difference around $150$--$200$~K. This indicates a temperature increasing with height at the sampled layers, which also is confirmed in Figs.~\ref{fig:CFs_shockwave}--\ref{fig:tg_CFs_t120}.
At the time of $t=120$~s, ($\Delta T_\mathrm{b} |_\text{B6}$) shows a negative value and ($\Delta T_\mathrm{b} |_\text{B3}$) a positive value, which is in line with what we expected from the specific example in Fig.~\ref{fig:tg_CFs_t120}.
The propagating shock front at $t_1=178$~s then gives rise to a rapid increase of sub-band difference at band~6 and a rapid decrease at band~3.
The fact that both of the receiver bands~3 and 6 probe the down-falling hot plasma at large altitudes prior to the second shock front (Fig.~\ref{fig:CFs_shockwave}), is indicated by the large positive sub-band differences (around $t=320$~s in Fig.~\ref{fig:Tb_ev_ex}b). The second propagating shock front therefore gives rise to \hea{negative continuum slopes} in both bands, similarly as for band~3 in the case of the first shock front. 
The shock signatures in the slope of the continuum thus differs depending on the physical conditions of the local atmosphere that the shock is facing and if the receiver band is probing plasma at low altitudes or down-falling plasma at larger altitudes prior to and after the sampling of the propagating shock front.
The time-derivatives of the sub-band differences (Fig.~\ref{fig:Tb_ev_ex}c) can in addition be used to reveal the shock fronts and in particular to more precisely see when the shock fronts start to be probed. For instance, at $t_1=178$~s, the peak at band~6 and the peak towards negative values at band~3, are prominent. 
Similarly, the start of probing the second shock front at $t_2=332$~s in band~6 and $t_3=345$~s at band~3 are indicated by the the steep peaks towards negative values.

There are however occasions showing an increase of brightness temperature that is not produced by a shock front. 
Around $t=275$~s, there is a quite rapid increase in $T_\mathrm{b}$ at band~3, which, as we can see in the contribution function and plasma temperature in Fig.~\ref{fig:CFs_shockwave}, is not due to sampling of a upwards propagating shock front, but rather by probing the down-falling plasma.
The sub-band difference shows however no indication of shocks and is steadily large between $400$ and $500$~K.
Between $t=270$~s and $t=310$~s, there is a steady increase in brightness temperature at band~6 on the same time as a slow increase of the sub-band difference is seen, which originates in a gradual shift from the lower component in the contribution, at heights around $0.7$~Mm to the upper component sampling the down-falling plasma.
This is difficult to interpret by looking only at the $T_\mathrm{b}$, but the evolution of $\Delta T_\mathrm{b} |_\text{B6}$ reveals the change from probing a layer with a lesser to a steeper gradient of the plasma temperature increasing with height.
The difference between the sub-bands thus gives a bit more information, that for instance can be used to distinguish the shock fronts from \hea{eventual other phenomena or changes of opacity at different layers leading to changes of formation height of the radiation at the sampled wavelengths.}

\begin{figure}[h]
\includegraphics[width=\columnwidth]{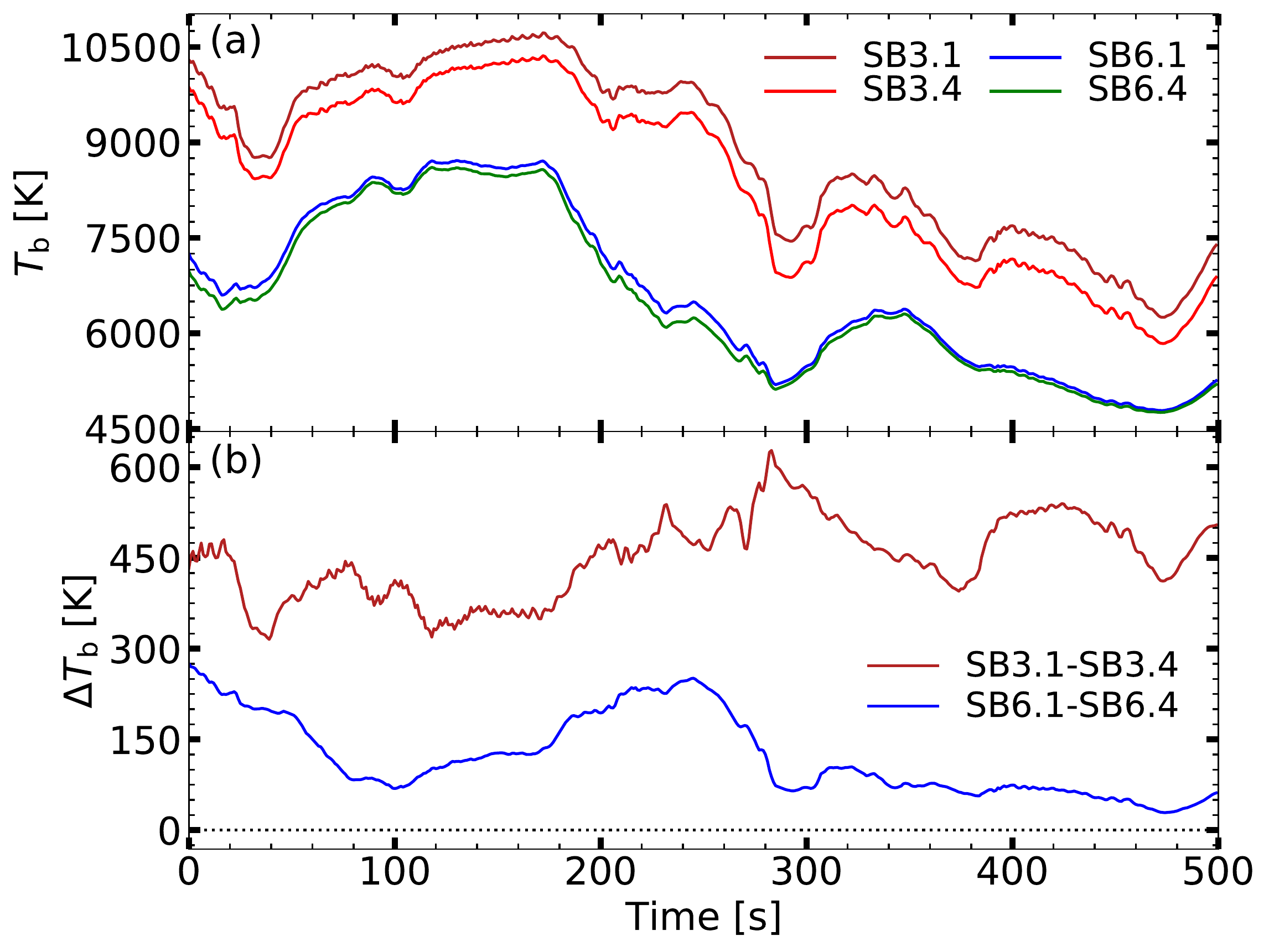}
\caption{Evolution of brightness temperatures and sub-band differences at location B in the network region between $t=0$--$500$~s. 
(a) The brightness temperatures of SB3.1, SB3.4, SB6.1 and SB6.4.
(b) Sub-band differences $\Delta T_\mathrm{b} |_\text{B3}$ ($T_\mathrm{b} |_\text{SB3.1} - T_\mathrm{b} |_\text{SB3.4}$) and $\Delta T_\mathrm{b} |_\text{B6}$ ($T_\mathrm{b} |_\text{SB6.1} - T_\mathrm{b} |_\text{SB6.4}$).
}
\label{fig:tb_tbdiff_NWex}
\end{figure}

\paragraph{Network region at location B.}

The brightness temperatures are on average higher in the network region than in the quiet Sun. See e.g., \cite{2021A&A...656A.68E} for statistics on the brightness temperatures at different regions in the \textit{Bifrost} model at the different ALMA receiver bands.
In Fig.~\ref{fig:tb_tbdiff_NWex}, the time evolution of the resulting brightness temperatures at location B in the network region (Fig.~\ref{fig:Tb_FOV}) is given.
This time-series thus corresponds to the contribution functions shown in Fig.~\ref{fig:CFs_network}.
Here, the brightness temperatures are significantly larger at band~3 than at band~6, by \hea{an average value of $2050$~K over the time-series}.
There are variations in brightness temperature over time also on the order of thousands of degrees with similar characteristics at both bands.
The differences in brightness temperature between the sub-bands are constantly positive at both bands~3 and 6, which is in line with what is seen at most locations in the central network region in the $\Delta T_\mathrm{b}$ maps (Fig.~\ref{fig:Tb_FOV}c and f). 
The positive sub-band differences reflect the increasing temperature with height at the sampled layers, which is indicated in Fig.~\ref{fig:CFs_network} and the larger $\Delta T_\mathrm{b}$ at band~3 than at band~6 shows the steeper gradient of the plasma temperature at the layer sampled by band~3 at larger altitude than the layer sampled by band~6. 
The variations in $\Delta T_\mathrm{b}$ are more prominent at band~3 than at band~6 (Fig.~\ref{fig:CFs_network}b) but are partly coupled to the variations in the brightness temperatures (Fig.~\ref{fig:CFs_network}a) at both bands.

\subsection{Sub-band differences with degraded angular resolution}\label{sec:degraded resolution}

\begin{figure*}[!htb]
\includegraphics[width=\textwidth]{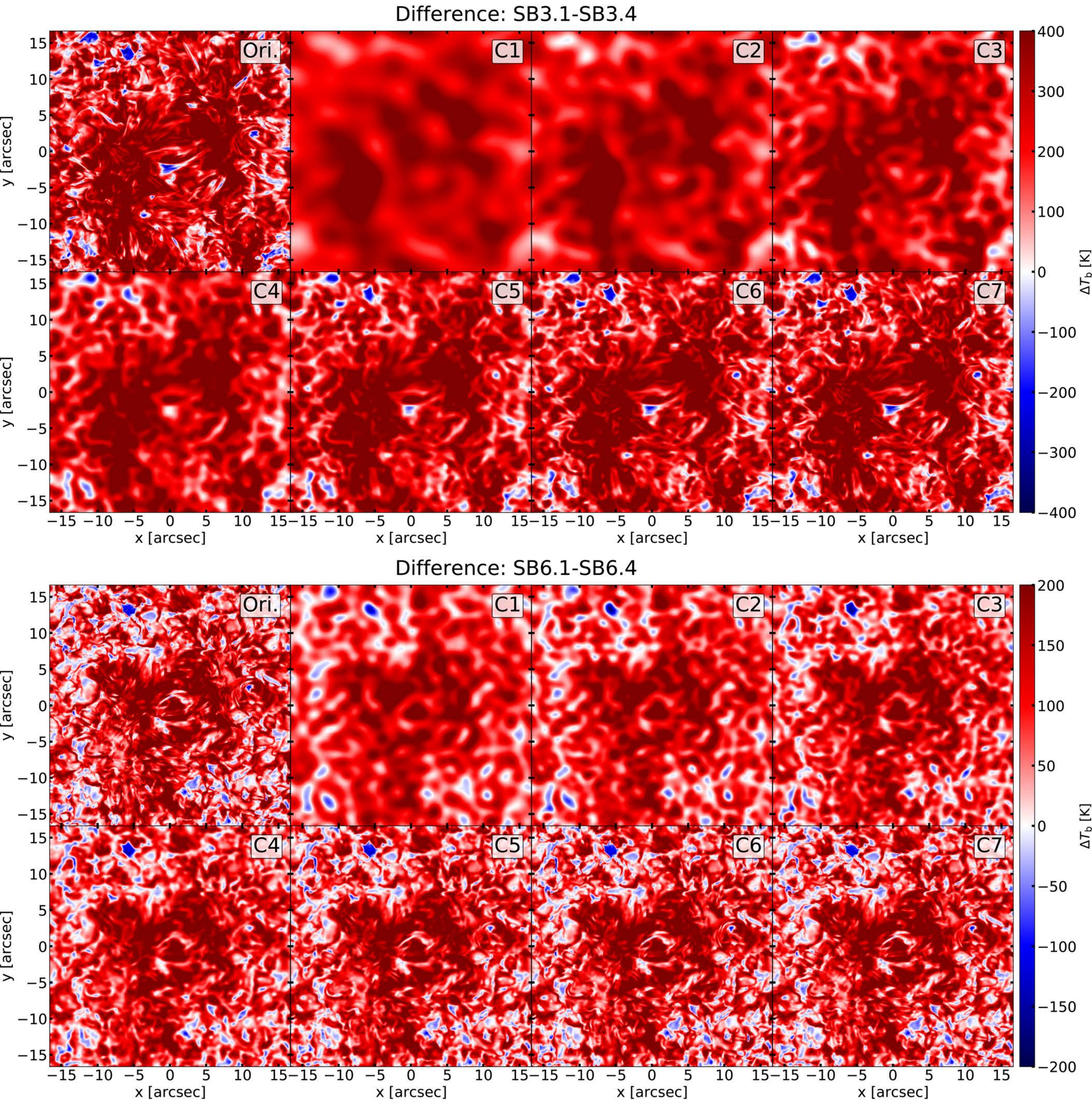}
\caption{\hea{The sub-band} brightness temperature difference \hea{at band~3}, $\Delta T_\mathrm{b} |_\text{B3}$, (top panels) and \hea{at band~6}, $\Delta T_\mathrm{b} |_\text{B6}$, (bottom panels) at original resolution of the numerical model and at the resolutions corresponding to array configurations C1 to C7 for respective receiver \ha{sub-band} (Table~\ref{tab:appendix_clean_beams}). \ha{It should be noted that the $T_\mathrm{b}$ maps are individually degraded to respective resolution before the differences are calculated.}}
\label{fig:sb_diff_FOV_b3}
\end{figure*}


The $T_\mathrm{b}$ differences between SB1 and SB4 for bands 3 and 6 over the entire FOV of a snapshot ($t=1600$~s) are presented in Fig.~\ref{fig:sb_diff_FOV_b3} at both original resolution and the resolutions corresponding to ALMA observations with array configurations C1-C7. 
\hea{Array configuration C1 is the most compact, resulting in an angular resolution down to about $3.4$~arcsec and $1.5$~arcsec at band~3 and band~6, respectively. The more extended configuration C7, which is not yet available for solar observing, would  result in a resolution down to about $0.23$~arcsec and $10.10$~arcsec at band~3 and band~6, respectively.} 
\hea{The analysis here extends to higher resolution than for the array configurations commissioned for solar observations (C4 at band~3 and C3 at band~6).}
The corresponding angular resolution for each array configuration and receiver sub-band are for reference listed in Table~\ref{tab:appendix_clean_beams}.

The variations of the $T_\mathrm{b}$ differences between the sub-bands are driven by the chromospheric small-scale dynamics (Sect.~\ref{sec:brightness temp.}). The typical spatial scales of the small-scale dynamics are on the same order or smaller than the resolution offered for solar ALMA observations \citep[see e.g.,][and references therein]{2021A&A...656A.68E}. The observability of the spatial fine structure of the sub-band differences is therefore very much dependent on the chosen resolution of the observations. 
Consequently, the small-scale structure in the maps of the sub-band difference becomes less pronounced at degraded resolution. 
As a result, 
\hea{negative continuum slopes}
occur less often or not at all 
in the degraded maps (Fig.~\ref{fig:sb_diff_FOV_b3}). 
The \hea{negative sub-band differences} is completely non-apparent in band~$3$ at the \hea{angular} resolutions corresponding to array configuration C1 and C2  (c.f.,~Table~\ref{tab:appendix_clean_beams}).
It is from this clear that the positive values of sub-band differences are the dominating feature.
A minimum resolution corresponding to array configuration C3 is necessary for the \hea{negative} sub-band \hea{difference} to be apparent in band~3.%
With a better angular resolution corresponding to the wider array configuration C4, which is available for Band~3 at the time of writing, more of the structures of the \hea{negative sub-band differences} are apparent, but preferably a resolution corresponding to C5 or higher should be used to see a larger correlation to the structures seen in the map at original resolution.
At band~6, the \hea{negative} sub-band differences ($\Delta T_\mathrm{b} |_\text{B6}$) are apparent at all offered resolutions (for solar observations), also corresponding to the most compact array configuration C1 (Fig.~\ref{fig:sb_diff_FOV_b3}).
This is a result of the significantly larger angular resolution at the wavelengths around $\sim1.25$~mm of band~6 compared to around $\sim3.0$~mm at band~3 (Table.~\ref{tab:almasubbands}), which preserves the observability of the features with strongest \hea{negative} sub-band \hea{difference}. 
However, the ($\Delta T_\mathrm{b} |_\text{B6}$) structures at sub-arcsec scales are only apparent at resolutions corresponding to array configuration C3 or higher. Although for the magnitudes of sub-band differences of structures at sub-arcsec to be well sampled, the resolution of C3 is not satisfying.
At locations where the sub-band difference is uniform over a large area in comparison to the clean beam, as for instance in the areas containing the magnetic foot points in the centre of the FOV (dark red patches in Fig.~\ref{fig:sb_diff_FOV_b3}), the magnitude and the large-scale structures of the sub-band differences remain also discernible at degraded resolution.

\begin{figure}[!htb]
\includegraphics[width=\columnwidth]{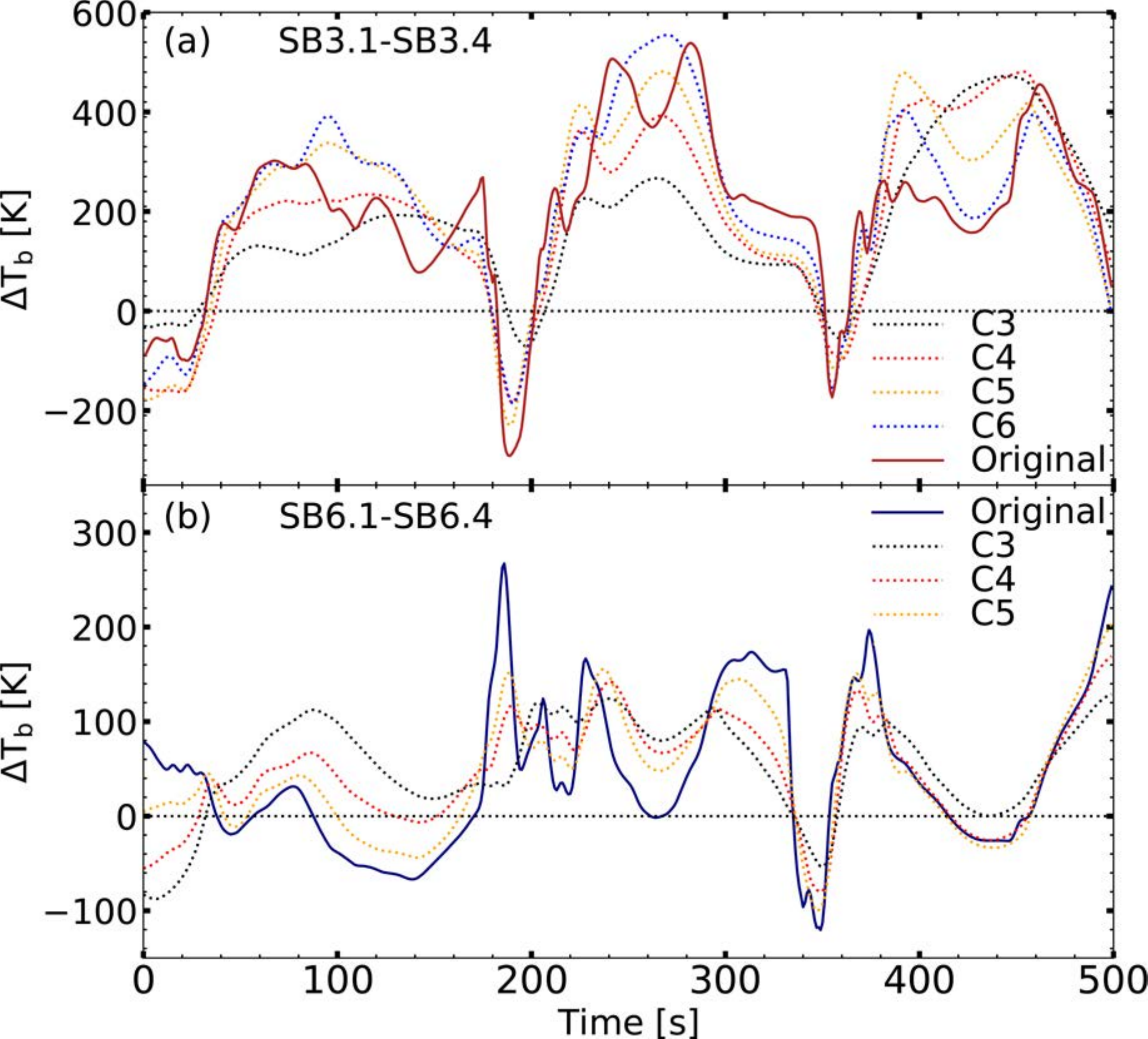}
\caption{
Time evolution of the sub-band $T_\mathrm{b}$ differences of the time-series at location A at the different angular resolutions. (a) The $T_\mathrm{b}$ difference between sub-bands 3.1 and 3.4 at original resolution (solid red line) and at degraded resolutions corresponding to array configurations C3, C4, C5 and C6 (colored dotted lines). 
(b) $T_\mathrm{b}$ difference between sub-bands 6.1 and 6.4 at original resolution (solid blue line) and at degraded resolutions corresponding to array configurations C3, C4 and C5 (colored dotted lines). The horizontal black dotted lines show the level of $\Delta$T$_\mathrm{b}=0$.
}
\label{fig:sb_diff_time_ev}
\end{figure}

The time series of the sub-band differences of bands 3 and 6, at location A  (Fig.~\ref{fig:Tb_FOV}) during the passage of the shock waves, in the maps with different angular resolutions are given in Fig.~\ref{fig:sb_diff_time_ev}. 
Here we see for example that during the first shock wave around $t=180$~s, the \hea{negative} sub-band \hea{difference} at band~3 of \hea{-291~K} at original resolution (Fig.~\ref{fig:sb_diff_time_ev}a), appears only as \hea{-73~K} at \hea{the resolution of the solar ALMA observations in this work (Sect.~\ref{sec:results - observations}; array configuration C3)}.
Similarly the large positive $\Delta T_\mathrm{b} |_\text{B3}$ values of up to $550$~K between $t=230$~s and $t=290$~s are shown only as a couple of hundreds K at the resolution of C3. 
Already at the resolution of C4 at band~3 (Table~\ref{tab:appendix_clean_beams}), the magnitudes of the sub-band differences are much more accurately sampled.
The magnitudes of the sub-band differences at band~6 are lesser than at band~3, which can lead to that weaker features are not seen at the degraded resolutions. 
For instance, a resolution corresponding to at least array configuration C5 is required to sample the \hea{negative} sub-band \hea{difference} at band~6 between $t=90$~s and $t=170$~s. 
It is thus expected to only see a \hea{negative continuum slope} at the angular resolution of the so-far offered observations with ALMA at some locations where the $\Delta T_\mathrm{b}$ is significantly large and the structure giving rise to \hea{negative continuum slopes} is large or at least at the same order of size as the resolution element of the observations.

However, the general features in the temporal evolution of the sub-band differences and the shock wave signatures are still preserved at the degraded resolutions corresponding to the higher resolutions currently offered for solar observations (i.e., minimum C3 at both bands).
Therefore an analysis of the sub-band differences in the observations provides much complementary information to that of the \hea{analysis of the full-band images}.

\subsection{Observational data}\label{sec:results - observations}

\begin{figure}[t!]
\includegraphics[width=\columnwidth]{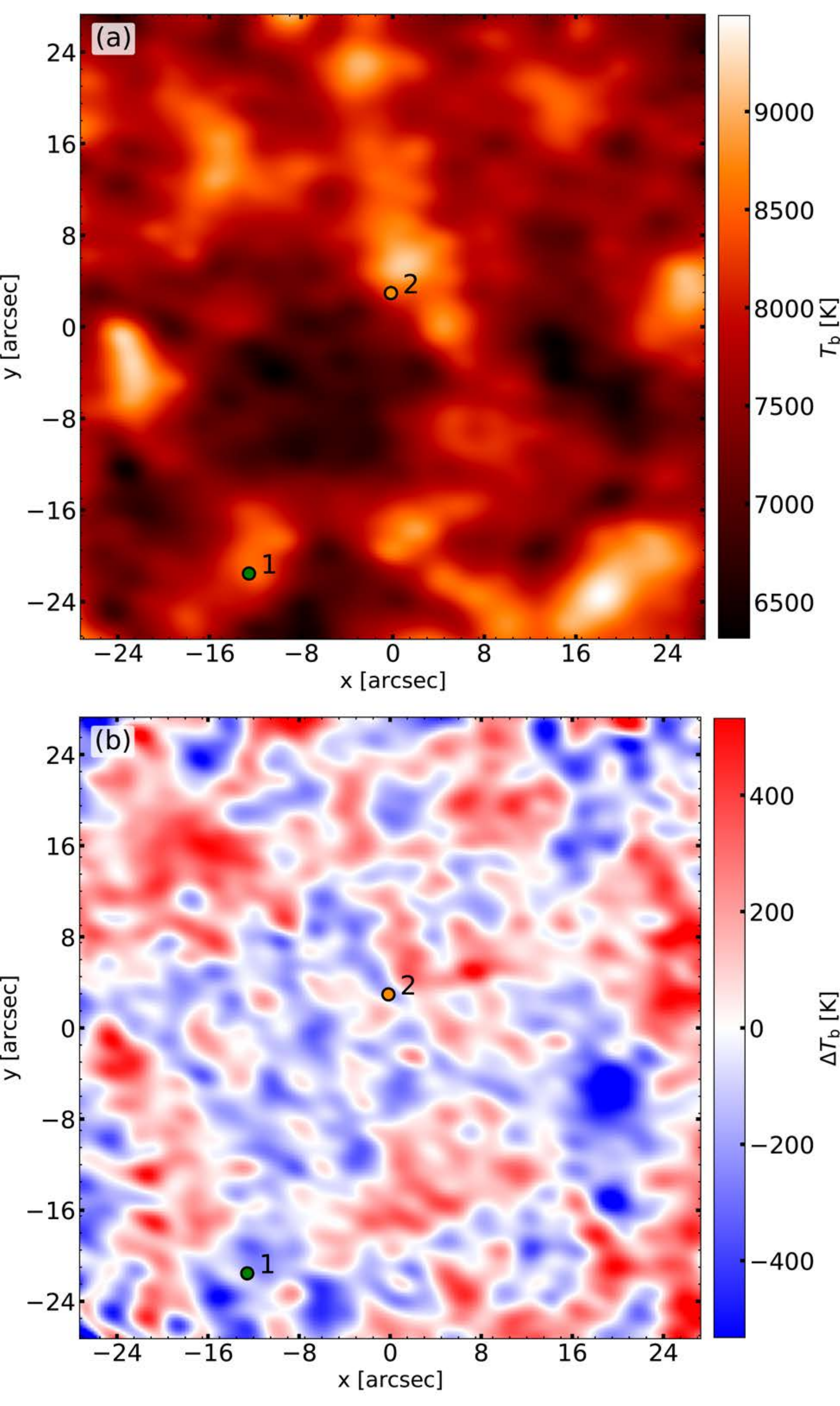}
\caption{Brightness temperature and sub-band brightness temperature difference of the FOV of the observational band~3 data at $t=290$~s.
(a) Brightness temperature of SB3.1. \hab{as constructed by adding the total power data to the interferometric data.} 
(b) Difference of brightness temperature between the sub-bands, $\Delta T_\mathrm{b} |_\text{B3} = T_\mathrm{b} |_\text{SB3.1} - T_\mathrm{b} |_\text{SB3.4}$. \hab{derived from only the interferometric data at each sub-band.} The span of $\Delta T_\mathrm{b} |_\text{B3}$ is between $-1150$~K and $670$~K, but the color scale was limited to $-500$~K to $500$~K to reveal the small-scale structures.  
The locations of the two selected brightening events are marked by the green and orange markers, respectively.}
\label{fig:obs_FOV}
\end{figure}

\begin{figure*}[t!]
\includegraphics[width=\textwidth]{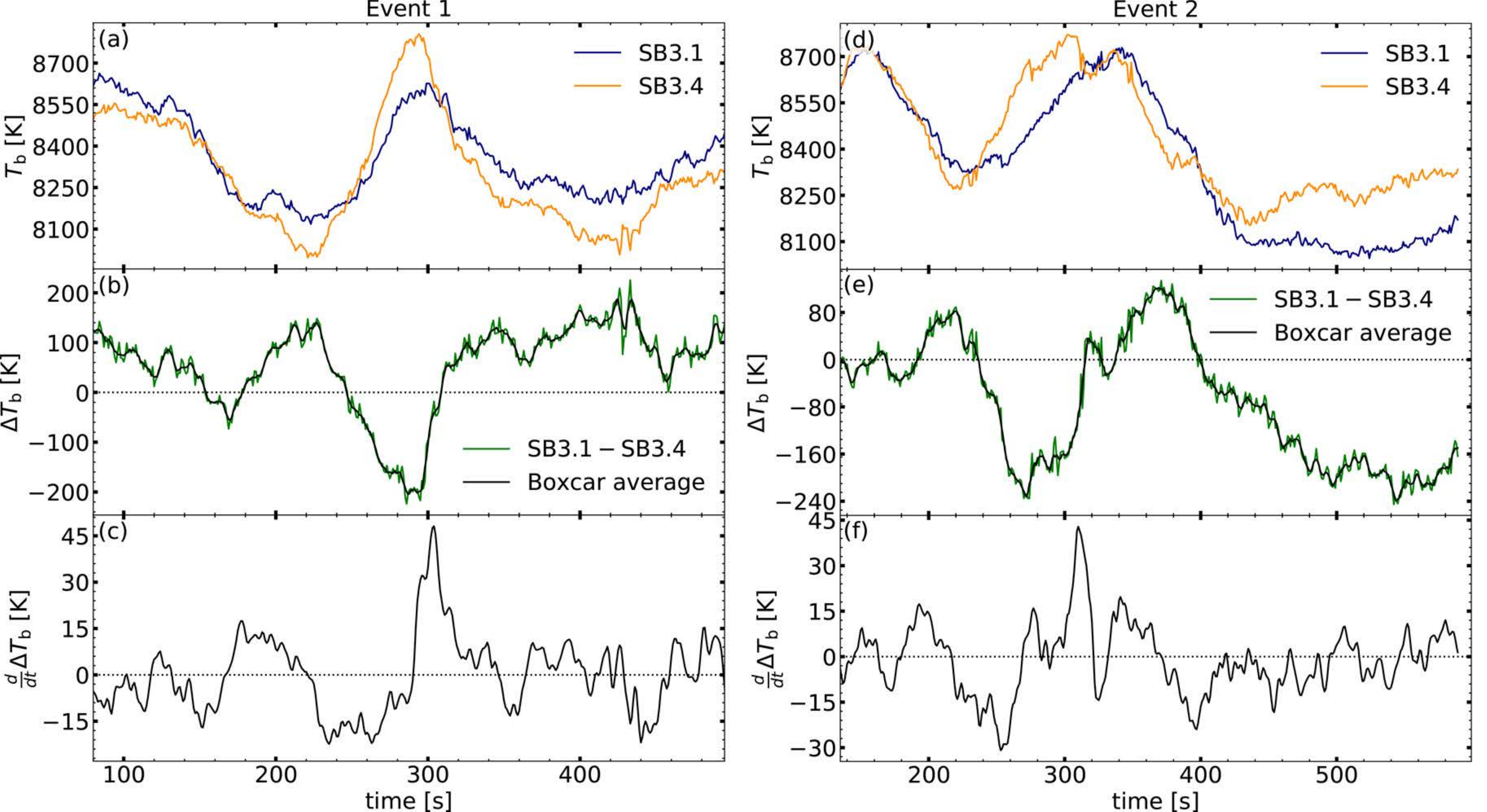}
\caption{Temporal evolution of the brightness temperatures at of the two selected brightening in the observational data. Left column: Event 1, Right column: Event 2. The locations of the events are indicated in the FOV of Fig.~\ref{fig:obs_FOV}.
(a,d) Brightness temperatures of SB3.1 and SB3.4. 
(b,e) Brightness temperature differences between the sub-bands $\Delta T_\mathrm{b} |_\text{B3}$. Green graph shows the original data and black curve a boxcar average over $10$~s. 
(c,f) The time derivative of the (boxcar averaged) sub-band differences. 
}
\label{fig:obs_ex1}
\end{figure*}

The brightness temperature of a snapshot of observational quiet Sun data at SB3.1 (Table~\ref{tab:almasubbands}) is shown in Fig.~\ref{fig:obs_FOV}a, for a FOV of approximately $50 \times 50$~arcsec. \hab{This image was made from adding the total power measurement to the interferometric data.}
\hab{Using only the interferometric data at each sub-band,} the sub-band difference,  $\Delta T_\mathrm{b} |_\text{B3} = T_\mathrm{b} |_\text{SB3.1} - T_\mathrm{b} |_\text{SB3.4}$ \ha{corresponding to} the same snapshot is \hab{acquired} (Fig.~\ref{fig:obs_FOV}b). \hab{The} color scale has been capped to $-500$~K to $500$~K 
\ha{from the span of} approximately $-1150$~K \ha{to} $670$~K.
\ha{The sub-band differences are strongly coupled to the small-scale features.}
\ha{The observational data features quiet Sun without any network patches \citep{2021RSTPA.379..174J}, which in the simulation is only comparable to the corners of the FOV (Fig.~\ref{fig:sb_diff_FOV_b3}).
The features with negative sub-band difference in the observation (Fig.~\ref{fig:obs_FOV}b) show typical scales up to about 5~arcsec, which agrees well with the scales of the few features that are seen in the corners of the simulation (Fig.~\ref{fig:sb_diff_FOV_b3}; note the different extents of the FOVs). 
A detailed study of the magnetic field strength and topology would need to be deployed for a global comparison of the sub-band differences of the observational data and the simulation. For now, we resort to analysing a couple of small-scale brightening events in the observational data.
The temporal evolution of the two events is shown in Fig.~\ref{fig:obs_ex1}.}
These events, 1 and 2, show an excess brightness temperature of approximately $450$~K and $400$~K, respectively at SB3.1, measured as the difference between the base temperature at the temporal local minimum and the peak temperature (see \cite{2020A&A...644A.152E} for definition and selection criteria of the base temperature). 
The temporal signatures and magnitudes of the brightening events are in line with what is expected for shocks propagating upwardly largely vertical through the chromosphere \citep{2021RSPTA.37900185E}, at the angular resolution of the observations \citep{2021A&A...656A.68E}. 
The differences between the sub-bands ($\Delta T_\mathrm{b} |_\text{B3}$) for the two events are given in Figs.~\ref{fig:obs_ex1}b and e, where a boxcar averages over $10$~s, removing potential high frequency noise, are also shown. $\Delta T_\mathrm{b} |_\text{B3}$ spans between approximately $-200$ and $200$~K during event~1 and between $-240$ and $100$~K during event~2. 
The temporal evolution of the $\Delta T_\mathrm{b} |_\text{B3}$ of both event 1 and 2 is consistent with what is seen for the synthetic observables of the shock wave event in the simulation (Fig.~\ref{fig:Tb_ev_ex}b), at band~3.
There is a rapid decrease of the sub-band difference at $t=225$~s for event~1 and at $t=220$~s for event~2, simultaneously with the start of the increase in brightness temperature (indicated by $t_1$ and $t_3$ in Fig.~\ref{fig:Tb_ev_ex} for the events in the simulation), \hea{which} is visible in the graph of the time-derivative of the 
sub-band differences in Fig.~\ref{fig:obs_ex1}c and f.
The $\Delta T_\mathrm{b} |_\text{B3}$ shows negative values during the increase of the brightness temperature (Fig.~\ref{fig:obs_ex1}a and d). 
During the end of the $T_\mathrm{b}$ peak, $\Delta T_\mathrm{b} |_\text{B3}$ increase abruptly, reflected by peaks in the time-derivatives (Fig.~\ref{fig:obs_ex1}c and f).

Correction for the degradation of the excess temperatures due to the angular \hea{resolution could be performed} for further in depth analysis of the brightening events \citep[][; in review]{2022A&A...submE}. However, in the current work, it is evident that the sub-band differences are large enough to be detected in the ALMA data and can be used for identifying shock signatures.

\section{Discussion}\label{sec:disc}

\subsection{Techniques for image reconstruction}

The synthetic brightness temperature maps based on the simulations are assumed to be equivalent to 
optimal observations with sampling on all scales and orientations (i.e., full Fourier sampling), enabling the study of sub-band differences and their connection to the atmospheric \hea{structures.} 
\hea{In} this work we perform analysis on the \hea{combined} brightness temperatures 
over each of the outermost sub-bands, i.e., SB1 and SB4 of \hea{bands~3 and 6  (Table~\ref{tab:almasubbands})}. 
\hea{Reconstructing the images} over a larger span of frequencies, for instance pairwise over both the neighbouring sub-bands SB1 and SB2 and compare that with \hea{the image of} both SB3 and SB4, would potentially improve the signal-to-noise ratio. This would however also diminish the magnitude of the sub-band differences and thus limiting this as a tool of measuring the slope of the continuum. 
The time integration used in the observations were one second which potentially could be extended to a few more seconds, in order to enlarge the signal-to-noise ratio.
However, this would potentially leave rapid small-scale dynamic features unresolved. 
The optimized integration over frequency and time for boosting the signal-to-noise ratio while preserving the important features thus depends on the specific science case and is worth to explore further.

As the sub-band differences are coupled to the small-scale dynamics, successful observations with more extended array configurations with resulting larger angular resolution would provide a very useful tool to probe the dynamic structures of the solar atmosphere. 
However, to what degree it is feasible to perform solar measurements at more extended array configurations, still needs to be determined. 

\ha{The detected signatures in the observational data agree well with what is seen in the simulation and this study of the sub-band differences serves as a proof of concept of the usage of sub-band differences for the analysis of solar ALMA observations. Nonetheless, further commissioning on calibrating the spectral domain of the ALMA receiver bands and deriving uncertainties on the sub-band differences is required.}

\subsection{Evaluation of the sub-band differences}

The sub-band brightness temperature difference (Eq.\ref{eq:SBdiffDef}) is dependent on the thermal structure of the atmosphere and the separation in formation height of the sub-bands.
\hea{Of the currently offered ALMA bands, band~$3$ shows the largest wavelength difference between the outermost sub-bands, SB1 and SB4 (Fig.~\ref{fig:formation_heights_FOV}), which results in an on average larger difference in formation height of the sub-bands, compared to the higher-numbered receiver bands (Table~\ref{tab:almasubbands}).}
\hea{However, even though the sub-band differences become more evident at longer wavelengths, the observational angular resolution goes down with increasing wavelength, which counteracts the ability to measure the sub-band differences. 
\hea{The} typical scales of dynamic features, e.g., such as the brightening events presented in \cite{2021A&A...656A.68E} are on the order of or even smaller than the angular resolutions currently offered for solar ALMA observations which limits the observability of them. 
}
Consequently, the small-scale structures of the sub-band differences are more prominent at the highest resolution offered for solar ALMA observations of band~6 than that of band~3 (Fig.~\ref{fig:sb_diff_FOV_b3}).

The \hea{sub-band} differences could be made larger if \hea{only} the \hea{intensities from}
\hea{the spectral channels in outermost part of the sub-bands, with maximum wavelength seperation,}
were \hea{combined in the imaging process, but this}
would however result in a lower \hea{signal-to-noise ratio.} 

\subsection{Atmospheric modeling and resulting radiation formation heights}

\hea{The resulting wavelength dependency of the height of formation of the radiation  depends on the physical processes included in the modelling of the atmosphere and the radiative transfer calculations.
\cite{2020ApJ...891L...8M} include the treatment of ion-neutral interactions (ambipolar diffusion) to their 2.5-dimensional \textit{Bifrost} simulations and report a similar height of formation of the synthetic observables at wavelengths of ALMA band~3 and band~6 in active region and network conditions. They also report that with accounting for non-equilibrium ionization rate of hydrogen and helium, the contribution functions are more concentrated to narrower height ranges for both ALMA bands, although there are occasions with two significantly strong components of the contribution function at different heights. 
However, in the current work at location B in the magnetic network region (Fig.~\ref{fig:CFs_network}), bands~3 and 6 also show to a large part a similar height of formation, in particular when the contribution function is concentrated and constrained to a rather narrow layer (up to $\sim t=200$~s), where after band~6 show two major components.

The brightness temperatures of the ALMA data are also useful for creating atmospheric models through inversion \citep{2018A&A...620A.124D}. It can however be difficult to create models with a good fit of the ALMA brightness temperatures to a specific layer in the atmosphere \citep{2022arXiv220508760H}, which could be explained by contributions from several major components at different heights (as shown in Figs.~\ref{fig:CFs_shockwave}-\ref{fig:CFs_network}). 
}

\section{Conclusion}\label{sec:conc}

\hea{The slope of the brightness temperature continuum provides} information on the gradient of plasma temperature of the local atmosphere at the sampled \hea{layers. A negative (positive) continuum slope indicates decreasing (increasing) temperature with increasing height.} 
\hea{The} time-dependent variations of the continuum slope reveal the evolution of the gradient of the plasma temperature at the sampled layers and can provide indications on the vertical structure of the atmosphere.
\hea{Propagating shocks give rise to large variations of the slope of the continuum, which therefore can be used for identifying and distinguishing shock waves from other events.}


The simulations show that there are often multiple major components at different heights contributing to the measured brightness temperatures in both bands~3 and 6, in particular in the quiet Sun region with propagating shocks, but also in the network region. 
As a consequence, the delay of signatures of upwardly propagating shock waves, between two sub-bands or different receiver bands, is not necessarily a measurement of the propagation speed of the shock front.
It \hea{can} rather \hea{be} a measurement of the difference in opacity of the upper component
between the two sub-bands or receiver bands. 
In the case where  
both bands sample \hea{a relaxed post-shock region that is perturbed by a new shock wave,} 
there is no \hea{significant} delay in the signature of the shock wave between bands~3 and 6.
Under quiet Sun conditions, band~3 more often shows multiple components at different heights than band~6.
This consequently makes the interpretation of the measured brightness temperatures \hea{in the quiet Sun} more complex at band~3 than at band~6.
In the network region, band~6 more often shows a second major (lower) component than band~3. 
\hea{There are also some certain locations in the network region of the simulation where both bands~3 and 6 display one single component of the contribution function, at similar heights.}
However, the sub-band differences can give an indication on if there are more than one major component contributing to the measured brightness temperature and their relative height of formation. It can therefore be very valuable to include the sub-band differences in the analysis of solar ALMA data.

The observability and utilization of the sub-band differences are limited by the angular resolution, because the scales of many dynamic features are on the same order of scales as the angular resolutions currently offered for solar ALMA observations.
While much of the small-scale structures remains unresolved, many features show a large sub-band difference that is apparent also at the angular resolution currently offered for solar observations.
Resulting from the high angular resolution, band~$6$ can therefore be used as a meaningful diagnostics tool for detecting and utilizing the sub-band differences.
Even though the numerical simulations indicate that occurrences of \hea{large} sub-band \hea{differences} are frequently occurring at band~3 as well, the lower angular resolution of the observations at band~3 limits their observability and \hea{negative continuum slopes} are observable at locations showing a very strong negative sub-band difference or over a well extended area of negative sub-band difference (large in comparison to the resolution element).

The results from the simulations are compared to observational data of a quiet Sun region, where the evolution of a couple of shock wave signatures are shown in detail. The variances of the sub-band differences and the \hea{negative continuum slopes} that are seen in connection to the shock signatures agrees with what is seen in the simulations, taking into account the limited spatial resolution of the observations.
The sub-band differences can thus be used as a method for detecting or differentiating shock wave signatures from other eventual transient events in the observational data.

\hea{With} the aid of the sub-band differences, further work can be developed towards in particular coupling the dynamical structures to the temperature stratification of the atmosphere at the sampled layers and \hea{identifying} the dynamical structures that gives rise to several major components of the contribution function.
\hea{In conclusion}, the slope of the brightness temperature continuum can be used as a diagnostics tool in the analysis of solar ALMA data and give invaluable information that contributes to the understanding of the dynamics that takes place and transport of energy in the solar atmosphere.


\section*{Acknowledgments}
This work was supported by the SolarALMA project, which has received funding from the European Research Council (ERC) under the European Union’s Horizon 2020 research and innovation programme (grant agreement No. 682462), and by the Research Council of Norway through its Centres of Excellence scheme, project number 262622, and through grants of computing time from the Programme for Supercomputing. HE was supported through the CHROMATIC project (2016.0019) funded by the Knut and Alice Wallenberg foundation.
This paper makes use of the following ALMA data: ADS/JAO.ALMA\#2017.1.00653.S. ALMA is a partnership of ESO (representing its member states), NSF (USA) and NINS (Japan), together with NRC(Canada), MOST and ASIAA (Taiwan), and KASI (Republic of Korea), in co-operation with the Republic of Chile. The Joint ALMA Observatory is operated by ESO, AUI/NRAO and NAOJ. We are grateful to the many colleagues who contributed to developing the solar observing modes for ALMA and for support from the ALMA Regional Centres.

\bibliographystyle{aa}
\bibliography{ms.bib}

\appendix
\section{Velocity distributions}\label{appendix:vertical_vel}
\begin{figure}[!htb]
\includegraphics[width=\columnwidth]{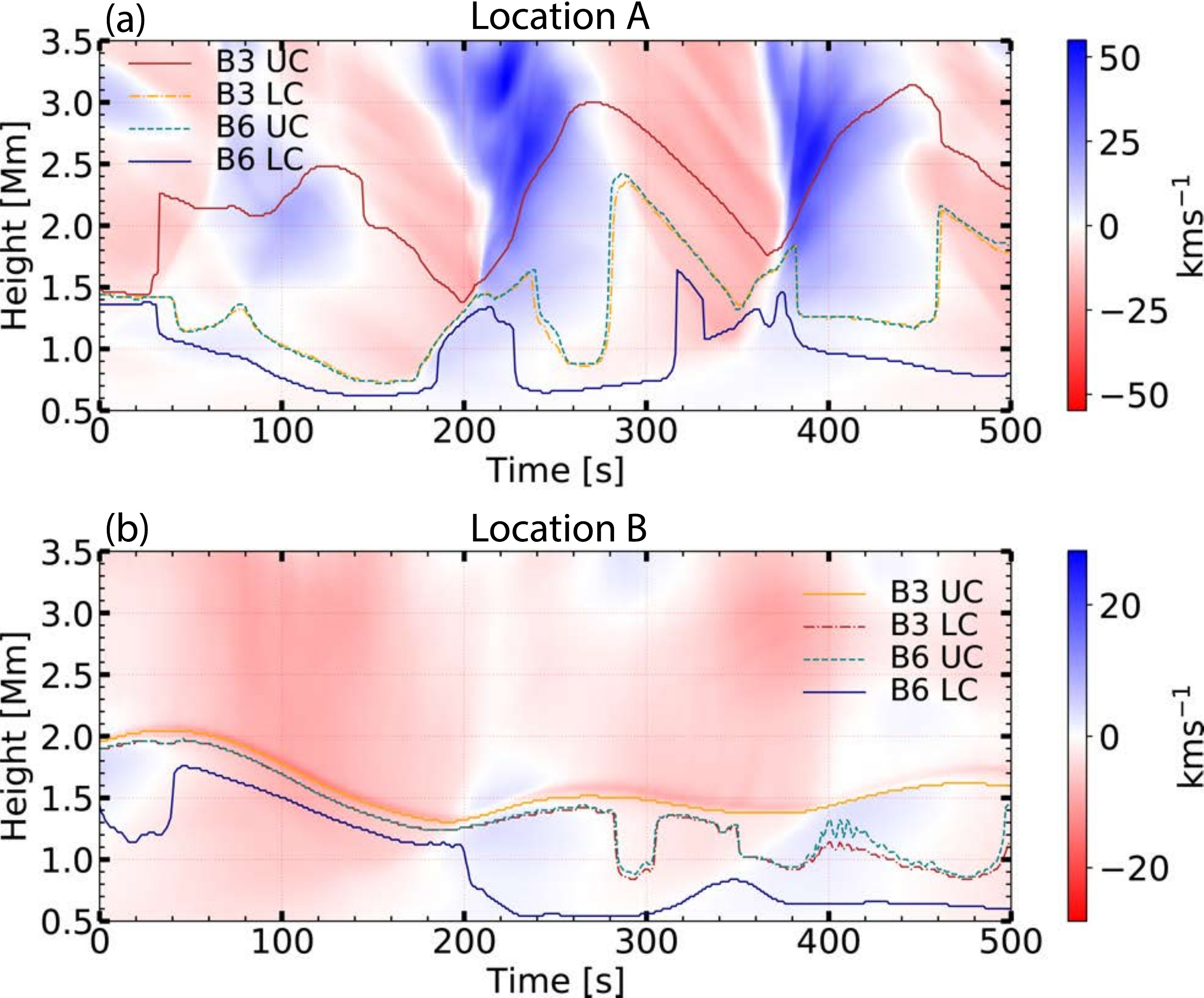}
\caption{Time evolution of the vertical velocities of the columns at location A during the propagation of shock waves (a) and at location B in the network region (b). Upwards motion, away from the photosphere, is indicated by negative velocity (blue color). The representative heights of formation of the lower and upper components of the observables for bands~3 and 6 are marked by the solid and dashed lines, respectively (see Sect.~\ref{sec:results} in the main text for details).
}
\label{fig:velocity_SIex}
\end{figure}

The evolution of the vertical velocity at the locations (A and B) of the two in-depth examples in the main text, are given in Fig.~\ref{fig:velocity_SIex}. The vertical velocity gives an indication on the up- and down-flow of the plasma at the different layers. The vertical velocity varies much in connection to the propagating shocks at location~A, while it remains lesser dynamic with dominating down-flow at location~B in the network region.

\section{Synthetic beam}
The synthetic beams used for downgrading the brightness temperature maps are given in Table~\ref{tab:appendix_clean_beams}.
As the numerical \textit{Bifrost} simulation does not represent a favoured orientation of the target on the Sun, all the setups were given the same position angle value of $80$ degrees, to make it possible to perform more one-to-one comparisons of the resulting observables from the different antenna configurations.

\begin{table*}[tbh!]
\caption{Clean beam parameters for SB3.1, SB3.4, SB6.1 and SB6.4 corresponding to that of ALMA array configurations $C1-C7$, that are used \hea{for the simulated data} in the current work. In the notation of the sub-band, they are preceded by the spectral band. The interferometric array configurations are the ones included in the simobserve tool \ha{in the \textit{Common Astronomy Software Applications package} \citep[CASA; v.6.1.0;][]{2007_McMullin_CASA}} for observational cycle 5, \hea{C1 being the most compact and C7 more extended. See \cite{2021A&A...656A.68E} for statistics on the baseline lengths. $^*$ The array configuration at this receiver band is currently not commissioned for solar ALMA observations.}}
\label{tab:appendix_clean_beams}
\centering
\begin{tabular}{ccccccccc}
\hline
Receiver&Sub-&ALMA Array&\multicolumn{3}{c}{Clean beam}\\
band&band&configuration &Major axis [arcsec] & Minor axis [arcsec] & Position angle [degrees] \\
\hline                
Band 3  & SB3.1&C1&   4.46 & 3.90 &  80\\
& SB3.1&C2&   3.28 & 2.83 &  80\\
& SB3.1&C3&   2.12 & 1.78 &  80\\
& SB3.1&C4&   1.33 & 1.18 &  80\\
& SB3.1&C5$^*$&   0.80 & 0.74 &  80\\
& SB3.1&C6$^*$&   0.51 & 0.40 &  80\\
& SB3.1&C7$^*$&   0.31 & 0.27 &  80\\
\hline                
& SB3.4&C1& 3.94 & 3.41 &   80\\
& SB3.4&C2& 2.89 & 2.48 &   80\\
& SB3.4&C3& 1.86 & 1.56 &   80\\
& SB3.4&C4& 1.16 & 1.04 &   80\\
& SB3.4&C5$^*$& 0.70 & 0.65 &   80\\
& SB3.4&C6$^*$& 0.45 & 0.35 &   80\\
& SB3.4&C7$^*$& 0.27 & 0.23 &   80\\
\hline
Band 6 & SB6.1&C1& 1.86 & 1.63 &  80\\
& SB6.1&C2& 1.38 & 1.17 &  80\\
& SB6.1&C3& 0.89 & 0.75 &  80\\
& SB6.1&C4$^*$& 0.56 & 0.50 &  80\\
& SB6.1&C5$^*$& 0.34 & 0.30 &  80\\
& SB6.1&C6$^*$& 0.22 & 0.16 &  80\\
& SB6.1&C7$^*$& 0.18 & 0.10 &  80\\
\hline                
& SB6.4&C1& 1.73 & 1.52 &   80\\
& SB6.4&C2& 1.29 & 1.09 &   80\\
& SB6.4&C3& 0.83 & 0.70 &   80\\
& SB6.4&C4$^*$& 0.51 & 0.47 &   80\\
& SB6.4&C5$^*$& 0.31 & 0.28 &   80\\
& SB6.4&C6$^*$& 0.19 & 0.15 &   80\\
& SB6.4&C7$^*$& 0.17 & 0.10 &   80\\
\hline
\end{tabular}
\end{table*}

\end{document}